
%


\input jnl_shore

\title Logarithmically slow domain growth in nonrandomly frustrated systems:
	Ising models with competing interactions

\vskip 0.2cm

\author Joel D. Shore, Mark Holzer, and James P. Sethna

\affil\lassp

\abstract
We study the growth (``coarsening'') of
domains following a quench from infinite temperature to a temperature $T$ below
the ordering transition.  The model we consider is an Ising ferromagnet on a
square or cubic lattice
with weak next--nearest--neighbor antiferromagnetic (AFM) bonds and
single--spin--flip dynamics.  The AFM bonds introduce
free energy barriers to coarsening and thus greatly slow the dynamics.
In two dimensions, the barriers are independent of the characteristic length
scale $L(t)$, and therefore the long time ($t \to \infty$) growth of $L(t)$
still obeys the standard $t^{1/2}$ law.
However, in three dimensions, a simple physical argument
suggests that for quenches below the corner rounding transition temperature,
$T_{CR}$, the barriers are proportional to
$L(t)$, and thus grow as the system coarsens. Quenches to
$T<T_{CR}$ should, therefore, lead to $L(t)\sim\log(t)$ at long times.

Our argument for logarithmic growth rests on the assertion that the mechanism
by which the system coarsens involves the creation of a step across a flat
interface, which below $T_{CR}$ costs a free energy proportional to its length.
We test this assertion numerically in two ways:  First,
we perform Monte Carlo simulations of the shrinking of a
cubic domain of up spins in a larger sea of down spins.  These simulations
show that, below $T_{CR}$, the time to shrink the domain grows
exponentially with the domain size $L$.  This confirms that the free energy
barrier, $F_B(L,T)$, to shrinking the domain is indeed proportional to $L$.
We find excellent agreement between our numerical data and an
approximate analytic expression for $F_B(L,T)$.
Second, to be sure that the coarsening system cannot somehow find paths around
these barriers,
we perform Monte Carlo simulations of the coarsening
process itself and find strong support for $L(t)\sim\log(t)$ at long times.
Above $T_{CR}$ the step free energy
vanishes and coarsening proceeds via the standard $t^{1/2}$ law. Thus,
the corner rounding transition marks the boundary between different growth laws
for coarsening in much the same way that the roughening transition separates
different regimes of crystal growth.

We also find logarithmic coarsening following a quench
in a two--dimensional ``tiling'' system which models the corner rounding
transition of a [111] interface in our three--dimensional model.
However, if instead of quenching, we cool the system
slowly at a constant rate $\Gamma$, we find the final length scale $L$
to have a power--law dependence on $1/\Gamma$, {\it i.e.},
$L \sim \Gamma^{-1/4}$, in accordance with a theoretical argument.

The predictions concerning the dynamics of the tiling model should in
principle be experimentally testable for a [111] interface of sodium chloride.

\pacs 64.60.Ht, 64.60.Cn, 05.50.+q, 64.70Pf

\endtopmatter

\chaptertitle I. \hskip 1em INTRODUCTION

It is known that in systems with externally imposed disorder (\ie randomness
in their Hamiltonians), the dynamics of ordering can be greatly slowed
because of free energy barriers which grow with the size of the
correlated regions.\refto{GrinsteinFernandez,FisherRFIM,Lai}  Motivated
by the slow dynamics present in glasses (discussed in Appendix~A),
we have searched for model systems in which such diverging barriers, and the
resulting slow dynamics, occur even {\it without} imposed
disorder.\refto{WhyNoDisorder} We have found\refto{Shore,JapanConferencePaper}
two closely related such models which we present in this paper.
In these models, the length scale with which the barriers diverge is not,
however, the equilibrium correlation length, but rather the characteristic
length scale $L(t)$ of the domains in a coarsening system.
By a coarsening system we mean one
which has been quenched from above to below its ordering
transition and is thus far out of equilibrium.  We will argue
that the barriers to coarsening grow linearly with $L(t)$, and that, as a
result, $L(t)$ grows only logarithmically with time $t$ following a deep
quench.  Before introducing the first model that we study, let us
review the relevent work on coarsening.

\beginsection A. Coarsening---A Brief Review\refto{MoreOnCoarsening}

When an Ising model is quenched from a
temperature well above its critical temperature, $T_c$, to a temperature below
$T_c$,  the system
finds itself in a disordered configuration at a temperature which favors
ordering.
At first, local regions of a few spins order.  However, since the ground
state is up--down degenerate, some local regions will order up and some down.
Thus, after a short time, one has a patchwork of up and down domains.
Since there is an interfacial free energy cost
associated with the total length (or, in three dimensions, the total surface
area) of the boundary between domains, this patchwork will evolve over time
so as to decrease the total boundary.  This process by which the system
orders over larger and larger length scales is called
``coarsening,'' although it is also commonly referred to as ``domain growth.''

Coarsening is ubiquitous in nature.  Examples include the coarsening
of foams (\eg bubbles in the ``head'' of beer), the coarsening of the
grains in a metal during the annealing process, the ordering of a binary alloy
following a quench from above to below its order--disorder transition, and
the phase separation of a binary fluid or alloy following a quench from the
one--phase to the two--phase region of its phase diagram.

In many ways, a coarsening system at late times is analogous to an equilibrium
system near a critical point.
For example, the coarsening system can be characterized by
a single ``characteristic length scale,'' $L(t)$ which generally grows with
some power of the time:
$$
        L(t) \sim t^n
        \ .
        \eqno(DivergingLengthScale)
$$

Since at late times, this length scale will be macroscopic,
one might expect that, just as in static critical phenomena, the exponent $n$
of the scaling will be independent of certain microscopic details of the system
(\ie of the lattice, the Hamiltonian, and the dynamics), and that there
will thus be only a few universality classes.
Two such universality classes have been delineated.
The first, with $n = 1/2$, is called
Lifshitz--Allen--Cahn,\refto{Lifshitz,AllenCahn} or ``curvature--driven,''
growth and occurs in many systems in which the order parameter is not
conserved by the dynamics.
Physical examples of this class include ordering in
binary alloys, grain growth in metals, and coarsening of foams.
The second, slower class, with $n = 1/3$, is called
Lifshitz--Slyozov\refto{LifshitzSlyozov} growth, and
occurs when the order parameter is conserved by the dynamics.
The primary example of this class is spinodal decomposition, the
process of phase separation in a binary fluid or alloy
quenched from its one--phase region (where the atoms mix) into the two--phase
region (where they separate).  That the order parameter in such
a system must be conserved follows from the fact that the total number of
atoms or molecules of each type is not changing with time.

Are these the only universality classes for coarsening?
An important lesson from static critical phenomena is that
universality is a subtle business.  That is, not all details of the system
are irrelevant, and distinguishing the relevant from the irrelevant
perturbations is not always obvious.  While it is very appealing to imagine
that all models fall into one of these two universality classes,
there is no {\it a priori} reason why we should expect this to be
the case.  In fact, as we will discuss further below,
it is already known that systems with imposed disorder, such as
the random field Ising model or an Ising model with dilute impurities,
will have logarithmically--slow coarsening (``$n = 0$'').

During the past decade, there have been various claims that, even in
certain models without randomness, coarsening would not obey the $t^{1/2}$ or
$t^{1/3}$ laws, but instead obey $L(t) \sim \log(t)$.
In particular, there were claims that in a $d-$dimensionsal system with
$Q \ge d + 1$
degenerate phases (\eg a Q-state Potts model),\refto{Lifshitz,LogPotts} and
in an Ising model with spin--exchange dynamics,\refto{LogIsingConserved}
the growth of $L(t)$ would be logarithmic, at least at low temperatures.
For a while, these claims could
not be disproved since the numerical evidence was
ambiguous due to long time transients and finite--size effects.  In particular,
the fact that domains in these coarsening models
were found to freeze entirely at zero temperature,\refto{PottsLattice}
and to grow only slowly at
very low temperatures, was taken by some as support for these claims.
However, large Monte Carlo simulations,
bolstered by more careful theoretical arguments, eventually showed that
the long time growth in these models obeys the naively--expected
power laws,\refto{NoLogPotts,VinalsandGrant,Huse,Amar,MCRGIsingConserved}
at least in two dimensions.

Two points should be drawn from this historical perspective.  First, one
must be careful when determining the growth law numerically.  The primary
reason that these controversies occurred in the first place is that numerical
simulations often give ambiguous results because they are hampered
by both finite--size effects and initial transients.  The former can become
a problem even when $L$ is only a fraction
of the system's linear dimension.\refto{VinalsandGrant}
The latter can affect results even out to quite late times, particularly
when there are energy barriers present which cause freezing after a quench
to $T = 0$.

Second, the relation between
such zero--temperature freezing and logarithmic growth has been widely
misunderstood.  Originally, there was the belief
that zero--temperature freezing would imply logarithmic coarsening at low $T$.
In light of the evidence showing this is not necessarily the case, the
pendulum of opinion seems to have swung the other way, with the assumption
made (at least implicitly) that zero--temperature freezing is always an
uninteresting phenomenon.\refto{Skepticism}  In fact, there seems to be
growing conviction that the standard $t^{1/2}$ and $t^{1/3}$ laws will hold
universally (at $T \ne 0$), independent of almost any details of the
Hamiltonian save randomness.  This conviction has
been stated, with varying degrees of generality, by several
authors.\refto{Universality}  It is important to note, however, that no one
has demonstrated the degree to which universality should apply.
The belief that it should apply so broadly is based mainly upon the
growing number of systems in which these laws hold and the
lack of any counterexamples.

It is Lai, Mazenko, and Valls\refto{Lai} who have stated clearly
when to expect logarithmic growth.  They
distinguish four different classes of systems on the basis of how the free
energy barriers to coarsening grow with $L(t)$.\refto{UniversalConfusion}
Emphasizing what we think is most fundamental, namely, the differential form
for the growth of $L(t)$, we present here (in a suitably modified form) the
delineation of these classes.
For definiteness, let's consider the differential equation for
the case of curvature--driven growth:
$$
	{\ddt L} = {a(L,T) \over L} \ . \eqno(DifferentialForm)
$$
(Completely analogous arguments follow for the Lifshitz--Slyozov case by
replacing the
$L$ in the denominator by $L^2$.) The four classes can then be distinguished by
the behavior of $a(L,T)$.

Class 1, in which $a(L,T)$ remains nonzero as $T \to 0$,
consists of those systems which do not have energy
barriers to coarsening and thus coarsen even at $T = 0$.  The canonical
example is the Ising model with spin--flip dynamics.
(This class, Lai \etal point out, is actually unphysical since all known
experimental examples of coarsening have elementary processes which involve
activation over barriers.)

The other three classes all freeze at zero temperature
(\ie $a \to 0$ as $T \to 0$).
Class 2 consists of systems whose energy barriers are
independent of $L$.  For example, if there is
only one such barrier height, $F_B$, then we can write
$a(L,T) = a_0 e^{-F_B/T}$.\refto{NoBoltzmann}
Integrating Eq.\ \(DifferentialForm) for this case gives
$$
	L(t) = \sqrt{ L_0^2 + a_0 t/\tau(T)}
	\ ,
	\eqno(Class2)
$$
where $\tau(T) = e^{F_B/T}$ is the characteristic time to surmount the
barriers.  In such systems, coarsening will be slow, with $L(t) \approx L_0$,
for times $t$ short compared to $\tau$; however, on
time scales long compared to $\tau$, activation over the barriers will be
common and one finds normal power law growth, $L(t) \sim t^{1/2}$ (albeit,
with a strongly temperature--dependent prefactor).

Class 3 and 4 systems are those in which the barriers grow with L.
Class 3 refers to the case where the barriers grow linearly with $L$,
while class 4 refers to the case where they grow like $L^m$ with $m \ne 1$.
Lai \etal note that all
known examples of these two classes are systems with disorder, \ie
randomness in their Hamiltonians.  An example of a system believed
to be class 3 is the random field Ising model,
while spin glasses and the Ising model with random quenched
impurities are thought to be class 4 systems.  For class 3, we can write
$a(L,T) = a_0 e^{-f_B L / T}$, where $f_B$ is a free energy barrier per unit
length.  Integrating Eq.\ \(DifferentialForm) then leads to a complicated
expression, which at long times goes asymptotically to the form
$$
	L(t) \sim {T \over{f_B}} \log(t)
	\ .
	\eqno(LogarithmicGrowth)
$$
Likewise, for class 4, $L(t) \sim [\log(t)]^{1/m}$ as $t \to \infty$.

\beginsection B.  Argument for logarithmic coarsening

What we will show, by example, in this paper is that there exist
models without randomness in their Hamiltonians which nonetheless have
free energy barriers that grow with the coarsening length (\ie are in
class 3 or 4), and thus have logarithmically--slow dynamics.
We now define a model and present the basic
argument for logarithmic coarsening in the three--dimensional (3D) version of
this model.

\beginsubsection 1. Introduction to the model

To study the dynamics of a statistical mechanical model, we must specify
two things:  the Hamiltonian and the dynamical rule for evolving the system.
The bulk of this paper (excluding Section~IV) will consider a system
whose Hamiltonian is simply
the ferromagnetic Ising model on a square or cubic lattice in $d = 2$ or $3$
dimensions, with dynamical frustration\refto{WhatFrustration}
added by introducing weak next--nearest--neighbor
antiferromagnetic (AFM) bonds.  The Hamiltonian is thus
$$
	H = -J_1 {\sum_{\rm NN} s_i s_j} + J_2 {\sum_{\rm NNN} s_i s_j}
	\ ,
	\eqno(hamiltonian)
$$
where the spins $s_i$ take on the values $+1$ and $-1$ (also called ``up''
and ``down'', respectively).  The first sum is over
all pairs of nearest--neighbors (NN) while the second is over all pairs of
next--nearest--neighbors (NNN).  We have chosen our sign convention so that
both $J_1$ and $J_2$ are positive when the NN bonds are
ferromagnetic and the NNN bonds are antiferromagnetic.  We will require
that $J_1 / J_2 > 2 (d - 1)$ so that the ground state is
ferromagnetic.\refto{Statics}

The dynamical evolution of the model will be governed by random--updating,
single--spin--flip dynamics.  This means that a spin is chosen at random
and flipped with a probability $P$ given by
$$
  P = {e^{-\Delta E/T}\over{1 + e^{-\Delta E/T}}} \ ,
  \eqno(Glauber)
$$
where $\Delta E$ is the change in energy that flipping this spin would
produce.\refto{ZeroTemperatureGlauber} The updating rule we use,
Eq.\ \(Glauber), is usually referred to as Glauber dynamics.\refto{Glauber}
Metropolis dynamics\refto{Metropolis} should give qualitatively similar
results.

\beginsubsection 2.
Expected behavior of the model in two and three dimensions

The following simple physical argument demonstrates that the NNN bonds
introduce energy barriers to domain coarsening.  Let us consider shrinking
a droplet of, say, up spins immersed in a sea of down spins.

First, we consider a square droplet in a two--dimensional (2D) system as
shown in \Fig{square_and_cube}(a). For simplicity, let us assume $T \ll J_1$ so
that only spin flips which do not raise the $J_1$ energy are accepted.
Without the NNN ($J_2$) bonds, such a square can shrink away without
the system having to cross any energy
barriers, since a corner flips for free, and then the edges can unravel for
free.  (This is why a nearest--neighbor Ising model will coarsen
even at $T=0$.)  However,
the NNN bonds introduce an energy barrier of $4 J_2$ to flipping a corner spin
(shaded dark gray)
since three of its four NNN spins (those spins diagonally away from it)
are pointing down.
Once the corner flips, the neighboring spins along the edge (shaded
light gray) can flip for free (and the final spin to flip along the
edge reduces the system's energy by $4 J_1$).  Therefore,
shrinking the square involves surmounting energy barriers of
height $F_B = 4 J_2$.  Note that the barriers to be crossed
are {\it independent} of the edge length $L$.  For time scales smaller than a
characteristic time $\tau = e^{4 J_2 / T}$, we expect
little domain coarsening to occur.  However, on time scales much
longer than $\tau$, such corner flips occur regularly and the $t^{1/2}$ law
should be observed.  (In the language of Lai, Mazenko, and Valls,\refto{Lai}
this is a class--2 system.)  Thus, although the dynamics is slowed, the
behavior of the growth law at asymptotically long times is not changed.
Because of this, the 2D case will be of little interest to us
except insofar as it provides a nice contrast to the 3D case.\refto{Pedagogy}

In three dimensions, the argument for the energy barriers to flipping a cube
of linear size $L$, shown in \Fig{square_and_cube}(b), is made analogously.
The energy barrier to flip a corner spin is now $12 J_2$ because nine of its
twelve NNN spins are pointing down (as can be seen by counting
the number of visible edges of the corner cube).
However, the important new feature which enters is that there is a
barrier of $4 J_2$ to flip each additional spin along the edge.
Since to flip an entire edge requires that
the corner spin flip and then each edge spin flip in turn,
the barriers add and the total barrier to remove an entire edge is
$$
	F_B(L) = 4 (L + 1) J_2
	\ .
	\eqno(OurBarriers)
$$
(This is simply the energy
difference between the initial state with the cubic domain and the
state in which all but one of the spins along an edge of the cube have flipped.
Once an entire edge has flipped,
there are other smaller energy barriers which must be crossed to further
shrink the cube.  However, in the low--$T$ or high--$L$ limit, the largest
of the barriers that must be crossed in sequence dominates the time to flip
the cube.)  We have already argued that barriers proportional to the length
scale $L(t)$ will yield logarithmically slow coarsening at long times.
A quick and dirty way to see this is to note that the time to flip a cube is
$t = \> \tau_0 \> e^{4 (L + 1) J_2 / T}$.  Simply inverting this
expression\refto{Lai} by solving for $L$ tells us that the coarsening length
grows as
$$
	L(t) \sim {T \over{4 J_2}} \log(t/\tau_0)
	\ .
	\eqno(OurLogarithmicGrowth)
$$

\beginsubsection 3. Outline of the paper

Our argument for logarithmic growth in the 3D model has
an appealing simplicity to it.  Unfortunately, however, it is only suggestive
and is far from a rigorous proof.  In Sections II and III, we will address the
two major objections which we can envision:

\item{(1)}  One can challenge our assertions concerning the barrier
to shrinking a cubic domain.  In particular, at nonzero temperatures,
one must consider not energy barriers but, rather, {\it free} energy
barriers.  The effects of entropy must be accounted for.  In Section~II,
we will find,
through both analytic work and simulations of shrinking cubes,
that our argument indeed breaks down above a temperature $T_{CR}$, which
we identify as the corner rounding transition
temperature previously studied in the context
of equilibrium crystal shapes.\refto{RottmanAndWortis}
Only below this temperature does the
free energy barrier to depin a step from an edge scale with the length of the
edge.  It is important to note that such a transition occurs only
when one has the discreteness introduced by the lattice.  Continuum models
will not have the pinned phase
(unless they explicitly contain a term to model this discreteness).
This is why the possibility of
logarithmically--slow coarsening is overlooked by the Lifshitz--Allen--Cahn
analysis.

\item{(2)}  One can challenge the claim that the barriers we have identified
to shrinking cubes imply logarithmic coarsening.  The jump in logic
from Eq.\ \(OurBarriers) to Eq.\ \(OurLogarithmicGrowth) is, indeed, a
large gap in our argument:  We have identified a special configuration
in which there are energy barriers which scale with the length scale $L$;
however, we have not shown that, during
the process of coarsening, the system will necessarily find itself in
configurations in which it will have to cross these barriers in order to
coarsen further.  It is conceivable that the system could find a way around
these barriers.  To construct a proof that the barriers must be crossed is
very difficult since it requires a detailed understanding
of the spin configurations which form in a quench.
Instead, in Section~III, we will be content to give some brief
arguments explaining why we think it is plausible that the barriers cannot be
avoided, and then ultimately, as is most common in this field, we will perform
numerical simulations of the quench to back up our arguments.

In Section~IV, we introduce what we dub the ``tiling model''.
This is a 2D model for a [111] interface in our 3D
Ising model.  This model is itself describable as an Ising model on a
triangular lattice, but has a more compelling visual representation
as a tiling of the plane by rhombi of three different orientations.  The
same basic arguments for logarithmic coarsening should hold in this model.
The model is a bit more obscure than, but
has two clear advantages over, the 3D model.  The first
is that, since it is two--dimensional, we can actually look at the
configurations which the system is getting stuck in.  The second is that
since the larger $J_1$ energy scale is removed naturally by introducing
a constraint on the allowed configurations, we can simulate the system
out to times which correspond to energy barriers much larger than
any elementary (single--spin--flip) energy barriers in the model.   This
makes it seem unlikely that the growth of $L(t)$ would resume power
law behavior at longer times.

In Section~V, we summarize, and briefly discuss some open questions to be
pursued.

Appendix~A presents a brief discussion of the slow dynamics in glasses,
our theoretical view on the matter,\refto{ToOurGlassPaper}
and how this view motivated our search
for logarithmically slow coarsening in systems without randomness.
Appendix~B gives a brief summary of how to most easily compute the
energy of configurations in our 3D model.   Appendix~C
contains a brief discussion on the implementation of the Monte Carlo algorithm.
A discussion of the scaling and anisotropy
of the correlation function for the coarsening
simulations of Section~III is presented in Appendix~D.

Finally, we note that some of the work presented here has appeared
elsewhere\refto{Shore, JapanConferencePaper} in abbreviated forms.  A somewhat
lengthier presentation than that given here can
be found in the first author's thesis.\refto{Thesis}

\chaptertitle II. \hskip 1em ANALYSIS OF SHRINKING CUBES

\beginsection A. Numerical simulations of shrinking squares and cubes

To investigate the free energy barrier for shrinking square or cubic droplets
at nonzero temperatures, we turn first to Monte Carlo simulations of this
process.

We start with the less interesting, 2D case.
\Fig{shrinking_squares.ps} shows the results of Monte Carlo simulations
for the time to shrink a square domain.
(Details of the Monte Carlo algorithm are given in
Appendix~C.)  Since this is an Arrhenius plot, activation over a constant
free energy barrier would give a straight line with the slope equal to the
barrier height.  Indeed, the
low--temperature Monte Carlo data can be fit well to a straight line.
The slope of the data is independent of the size of the
domain, implying a free energy barrier independent of domain size.
Furthermore, the form
$$
	t = \tau_0(L) e^{4 J_2/T}
       \ ,
       \eqno (Shrink2d)
$$
with $\tau_0(L)$ as a free parameter gives fine fits to the data, thus
demonstrating that the free energy barrier is $4 J_2$, as expected.

\Fig{shrinking_cubes_0_full.ps} shows the time to
flip the first edge of a cubic domain in three
dimensions.\refto{TimeToShrink}
Again the low--temperature data is quite
straight on this Arrhenius plot, thus indicating that shrinking a cube is
also an activated process.  However, in contrast to two dimensions,
here the slope, and thus the free energy barrier,
is clearly an increasing function of the size of the domain.
In \Fig{shrinking_cubes_0_full.ps}, we have tried a fit to the expected form
$$
        t = \tau_0(L) e^{4 J_2 (L+1)/T}
       \ ,
       \eqno (Shrink3d)
$$
with $\tau_0(L)$ as a free parameter.  For $L = 4$,
the fit appears quite good at low temperatures, but for $L = 6$ it is
marginal, and for $L = 8$ and especially $L = 12$ the fit is clearly
inadequate (down to the lowest temperatures we can reach).

The lack of agreement with the predicted zero--temperature barrier is
not too surprising since we expect
the free energy barrier to decrease as the temperature is raised.
What may seem surprising at first is that the slope of the data is larger
than the predicted slope of $4 J_2 (L + 1)$, tempting us to
conclude that the free energy barrier is in fact {\it larger} than predicted.
That this is not so can be seen by considering the
general form for Arrhenius activation over a free energy barrier $F_B(L,T)$:
$$
	t = \tau_0(L) e^{F_B(L,T)/T}
       \ .
       \eqno (Shrink3dForm)
$$
If we take the logarithm of
this expression and differentiate with respect to $1/T$, we get that the
slope on an Arrhenius plot is
$$
        {d[\log(t)]\over{d[1/T]}}  = F_B - T {dF_B\over{dT}}
        \ .
	\eqno (SlopeOnArrheniusPlot)
$$
Since $dF_B/dT$ will be negative, the slope on the Arrhenius plot is greater
than the free energy barrier.  Physically this is because if the free energy
barrier increases as we lower the temperature, then the resulting rise in the
activation time $t$ is due partly to the direct consequence of the
decrease in the temperature and partly to the fact that the barrier itself has
increased.  In fact, the righthand side of
Eq.\ \(SlopeOnArrheniusPlot) will be greater than the zero temperature
barrier, $F_B(L,T=0)$, if $F_B(L,T)$ is concave down over the region between
$0$ and $T$.  We expect such negative concavity on general grounds
since we anticipate that $dF_B/dT = 0$ at $T = 0$
and that it becomes negative for $T > 0$.  Thus
a steeper slope than $4 J_2 (L + 1)$ is perfectly consistent
with the notion that the free energy barrier $F_B(L,T)$
is decreasing with temperature.

Returning to \Fig{shrinking_cubes_0_full.ps}, we see that there are
more features to explain at higher temperatures.  For temperatures
above $T \approx 6 J_2$, the data appear to enter a regime where the
free energy barrier becomes roughly independent of size.
In Subsection~D, we will explain that the temperature marking
this change from an $L$--dependent barrier to an $L$--independent barrier
occurs at the corner rounding temperature,
$T_{CR}$, of the associated equilibrium crystal shape
problem.\refto{RottmanAndWortis}
This is then the temperature at which we expect
our argument for logarithmic coarsening to break down.
(The higher temperature
of $T \approx 100 J_2$, or more suggestively, $T \approx J_1$,
at which the barrier appears to vanish altogether
is most likely merely a quirk of our criterion for determining when the first
edge of the cube has disappeared.  This criterion becomes suspect once
$T/J_1$ is large enough for there to be a significant probability of
thermal fluctuations in equilibrium.)
First, however, let us study in more detail the
behavior of the time to flip an edge of a cubic domain
in the low temperature regime.

\beginsection B. Analytic calculation of the time to flip an edge of a cube

The goal of this Subsection is to derive an analytical expression to
match the simulation results for flipping the first edge of a cube
(\Fig{shrinking_cubes_0_full.ps}).
We will work in the limit $J_1/J_2 \to \infty$.  This is justified
by the insensitivity of the simulation results to $J_1/J_2$ in the low
temperature region where our expression will be valid.

To get the total rate for flipping a cube edge, we must sum over the rates for
all possible paths in configuration space which go from the state of a complete
cube to one with spins along an entire edge (and perhaps other spins) flipped.
The rate along any such path is just $\Gamma_i = {\Gamma_0}_i \> e^{-E_i/T}$.
Here, $E_i$ is the energy barrier (the maximum energy above the
initial configuration) along the given path and ${\Gamma_0}_i$ is a prefactor
which we expect to have some weak temperature dependence.  (In particular,
${\Gamma_0}_i$ should go to a nonzero constant in the limit $T \to 0$.)

At this point, it is useful to make an analogy with equilibrium statistical
mechanics.  The analogy is made by defining a barrier partition function
$Z_B$ as
$$
    Z_B \equiv  {\sum_{i} \exp(-E_i/T)}
       \ ,
       \eqno(PartitionFunctionDefinition)
$$
where the sum ranges over the barrier configurations
[See \Fig{cubes_galore}], {\it not} all paths.
The free energy barrier $F_B$ is then given by
$$
	F_B \equiv -T \log(Z_B)
       \ .
       \eqno(FreeEnergyBarrierDefinition)
$$
In terms of $F_B$, the time to flip the spins along the edge of a cube
can now be written as
$$
        t \equiv {1\over{\sum_{i} \Gamma_i}}
	\; = \; {1\over{\overline \Gamma_0}} \> \exp(F_B/T)
       \ ,
       \eqno(TimeFromFreeEnergyBarrier)
$$
where
$$
	{\overline \Gamma_0} \equiv {{\sum_i {\Gamma_0}_i
		\exp(-E_i/T)}\over{Z_B}}
       \ .
       \eqno(Gamma0Definition)
$$

Unfortunately, computing the prefactor ${\Gamma_0}_i$ for each barrier
configuration is prohibitive.  Therefore, we will make the following
approximation:  We set ${\overline \Gamma_0}$ equal to an estimate
of the prefactor ${\Gamma_0}_i$ for the {\it lowest}
energy barrier.  This estimate, in turn, is obtained by considering
only flips of spins along an edge of the cube, occurring sequentially
starting from one corner.  The problem can then be formulated as
a one--dimensional master equation, for which an exact solution can be
obtained in the limit $L \to \infty$.\refto{Thesis}  The expression for
${\overline \Gamma_0}$ thus found is
$$
    {\overline \Gamma_0}\; =\;
		 {12\; (1-e^{-4 J_2/T})^2\over{1-e^{-4 J_2/T}+e^{-12 J_2/T}}}
       \ .
       \eqno (Gamma0Average)
$$
Since the second lowest energy barrier is $8 J_2$ higher than
the lowest, using this expression
for $\overline \Gamma_0$ introduces an error on the order of
$e^{-8 J_2/T}$ into our computation of $t$.

We will now present an approximate calculation
for the free energy barrier $F_B(T)$.  Although the technique is similar to a
low temperature expansion for the equilibrium free energy,
two points should be kept
in mind.  The first is that we want to include only ``barrier configurations.''
That is, for each path in configuration space, we are interested only in
the configuration with the maximum energy along that path.
\Fig{cubes_galore}(a)--(d)
are examples of configurations that should be summed over, while
\Fig{cubes_galore}(e) is an example of one that should not
because a path in configuration space passing through such a configuration
must first pass through a higher energy configuration.

The second point is that we are not only interested in the thermodynamic limit
($L \to \infty$), because we wish to compare the resulting analytic expression
to numerical data for quite small
cubes.  In Subsection~C, we will find that the expressions derived here
simplify considerably in the thermodynamic limit.  That limit
is also much more forgiving:  Many approximations (concerning which
configurations to include) are irrelevant in
the thermodynamic limit, but do make a significant difference for the
cube sizes ($4 \le L \le 24$) in the Monte Carlo simulation.

Before proceeding with the calculation, let's summarize the
approximations which will be made in computing $F_B$:

\item{1.} We will continue to
work in the $J_1/J_2 \to \infty$ limit.

\item{2.} We will allow only the spins in one layer of the cube
to flip. \Fig{cubes_galore}(f)
shows an example of a neglected configuration.  Since the number of neglected
configurations is generally smaller (\ie grows with a lower power of $L$)
than the number of those with the
same energy that are included, this should be a fairly benign
approximation even for finite cubes.  In the thermodynamic limit
(and $J_1/J_2 \to \infty$), the free energy barrier (per unit edge length)
computed in this way becomes exact.

\item{3.} We will enumerate the configurations approximately, \eg by
extending certain sums to infinity even though they would
terminate for a finite cube.  These again are the sorts of approximations
which are irrelevant in the thermodynamic limit, but which can make
a difference for finite cubes.

\noindent Finally, we remind the reader that
in calculating the activation time $t$ from $F_B$,
we will approximate ${\overline \Gamma_0}$ as given
by Eq.\ \(Gamma0Average).

We proceed with the calculation as follows:
First, we will consider only configurations where the edge is being
``eaten away'' from just one of the corners
[\ie neglecting configurations like \Fig{cubes_galore}(d), which have energies
at least $8 J_2$ higher than the configuration in  \Fig{cubes_galore}(a)].
What we are then studying is simply the energies for various configurations
of a step across the face of the cube.
These energies can easily be computed using the two rules given in Appendix~B,
which
associate an energy with each plaquette (unit area) of interface and each
bend in the interface.
Since our approximations have
reduced the model to a simple one--dimensional solid--on-solid (SOS) model,
the energy of any configuration is specified
by a series of nonnegative integers $m_i$ giving the difference in the step
height between successive columns, as shown in \Fig{cube_for_energy.ps}.
The energy associated with each such column is simply
$$
  	E(m_i) = \cases{0, &if\ \ $m_i = 0$\ ;\cr
  		4 J_2 (m_i+1), &if\ \ $m_i > 0$\ .\cr}
	\eqno(EnergyOfConfig1)
$$
If we ignore the finite length of the cube in the vertical
direction,\refto{HighOrder} then we can allow each $m_i$
to range over all nonnegative integers (independent of
each other).  The partition function can thus be written
as\refto{GrubbyDetails}
$$
        Z_B = e^{-4J_2(L+1)/T} \; \bigl (2 Z_1^{L-2} - 1 \bigr )
        \ ,
        \eqno(FreeEnergyBarrier2)
$$
where $Z_1$ is the partition function for a single column:
$$
        Z_1 = { \sum_{m=0}^\infty e^{-E(m)/T}} \; = \; 1 + y
            \ , \eqno(FreeEnergyBarrier3)
$$
where
$$
        y \equiv {e^{-8J_2/T}\over{1 - e^{-4J_2/T}}}
        \ .
        \eqno(y)
$$
The free energy barrier (within the approximation that the edge is eaten away
from only one of the corners) is then
$$
	F_B = \;4\>J_2\>(L+1)\; -\; T\>
		\log \Bigl [2 {(1 + y)}^{L-2} - 1 \Bigr ]
       \ . \eqno(FreeEnergyBarrier4)
$$

If we now relax the assumption that the edge is eaten away from only one
corner and, instead, allow it to be eaten away from both corners [as in
\Fig{cubes_galore}(d)], the
partition function can be written as
$$
        Z_B = e^{-4J_2(L+1)/T} \Biggl [\bigl (2 Z_1^{L-2} - 1 \bigr )
	\> + \> y \> { \sum_{{\cal L}=0}^{L-3} \>
	\bigl (2 Z_1^{\cal L} - 1\bigr ) \bigl (2 Z_1^{L - 3 - {\cal L}}
	- 1 \bigr )} \Biggr ]
       \ .
       \eqno(FreeEnergyBarrier5)
$$
(Note that this {\it does} include the possibility that the two corners
are peeling away two different faces, as occurs in \Fig{cubes_galore}(d)
where the top face is peeling away from the right corner while the front
face is peeling away from the left corner.  Such configurations are important
to count since they have the same energy as the corresponding configurations
in which the same face is
being peeled away from both corners.)  Summing the geometric series, we
obtain our final result:
$$
        F_B =\;\; 4J_2(L+1) - T \log\Biggl [3 - 2 (1+y)^{L-2}
        +\; (L-2)\; y\;\Bigl (1 + 4 (1+y)^{L-3}\Bigr ) \Biggr ]
       \ ,
       \eqno(FreeEnergyBarrier7)
$$
where $y$ is given by Eq.\ \(y).

Eq.\ \(FreeEnergyBarrier7)
is our best approximation for the free energy barrier to flip the
first edge of a cubic droplet.  Substituting this into
Eq.\ \(TimeFromFreeEnergyBarrier) gives us
the estimate of the time to flip the first edge of a cube,
shown by the solid curves in \Fig{shrinking_cubes_2.ps}.
Clearly they are in excellent
agreement with the Monte Carlo data at low temperatures.

\beginsection C. The dynamical transition temperature

Now, we will take the thermodynamic limit $L \to \infty$
in order to calculate the free energy barrier per unit edge
length, $f_B$, to flipping the spins on an entire edge of a very large cube.
At the temperature, $T_{CR}$, where $f_B \to 0$,
the time to flip a cubic domain will no
longer depend exponentially on its edge length. Thus, $T_{CR}$ should mark
the transition in the coarsening
dynamics from logarithmic to $t^{1/2}$ coarsening.
As will be explained in Subsection~D, the
temperature $T_{CR}$ we obtain is also precisely the
temperature for an interfacial phase transition known as the
corner rounding
transition (hence our use of the subscript ``CR'').\refto{RottmanAndWortis}

We start with our best approximation to $F_B$, Eq.\ \(FreeEnergyBarrier7).
In the thermodynamic limit, this expression simplifies considerably
and we find that the free energy barrier
per unit edge length becomes
$$
        f_B \equiv {\lim_{L \to \infty}} {F_B\over{L}}
	\; = \; 4 J_2 - T \log \biggl (1 + {e^{-8J_2/T}\over{1 -
	e^{-4J_2/T}}} \biggr )
       \ .
       \eqno(FreeEnergyPerLength2)
$$
A plot of $f_B(T)$ is shown in
\Fig{free_energy_per_unit_length}.\refto{MightExpect}
The temperature $T_{CR}$ at which $f_B = 0$ is given by the cubic equation
$ x^3 - x^2 + 2x - 1 = 0$, where $x \equiv e^{-4J_2/T_{CR}}$.
This cubic has one real root which yields the result
$$
        T_{CR} = {-4 J_2\over{\log \biggl ({1\over{3}} -
			{5\over{9\beta^{1/3}}}
                        + \beta^{1/3} \biggr )}} \; \approx \; 7.1124\ldots J_2
        \ ,
        \eqno(TCR)
$$
where $\beta \equiv {1\over{6}} \Biggl( {11\over{9}} + \sqrt{{23\over{3}}}
\Biggr )$.\refto{CruderEstimate}
Since all the approximations we made in deriving $f_B$ are irrelevant
in the thermodynamic limit, our expression for the dynamical transition
temperature should be exact in the limit ${J_1/J_2 \to \infty}$.

Finally, we might ask about corrections to $T_{CR}$ for finite $J_1/J_2$.
Qualitatively, we expect $T_{CR}$ will drop with decreasing
$J_1/J_2$ and go to zero at $J_1/J_2 = 4$, where the ferromagnetic ground
state becomes unstable.    To obtain a good estimate for $T_{CR}$
near $J_1/J_2 = 4$ would require including configurations involving multiple
layers (see Ref.\ \cite{Thesis}, Appendix B), and thus a
much more concerted attack on the problem.  We can, however, obtain an
estimate of the importance of the corrections to our estimate of
$T_{CR}$ for finite $J_1/J_2$ by noting
that they will come in with a factor like
$e^{-4 (J_1- 4J_2)/T_{CR}}$.  This suggests $T_{CR}$ remains fairly
close to its $J_1/J_2 \to \infty$ value until $J_1/J_2$ gets quite close to 4.
For example, $J_1/J_2 = 12$ would yield a correction to $T_{CR}$ on the order
of $1\%$ or less.

We have already made several allusions to the relation between the temperature
at which the free energy barrier per unit length goes to zero and an
equilibrium interfacial transition known as the corner rounding transition.
In the following Subsection, we will give a brief discussion of the
interfacial
phase transitions in this model and then finally make explicit the relationship
between the coarsening dynamics and the corner rounding transition.

\beginsection D. Connection with equilibrium crystal shapes

The nearest--neighbor Ising model on a cubic
lattice has a roughening transition at $T_R \approx
2.45 J_1$.\refto{Burton,Weeks,vanBeijeren,HolzerandWortis}
Below this transition, interfaces in equilibrium are macroscopically
smooth in the sense that the variance of the fluctuations in the height of
the interface remains bounded as the system size is taken to infinity.
Above $T_R$, the interface is rough and the
variance diverges as the logarithm of the system size.

The roughening transition can be thought of in two ways:\refto{vanBeijeren}
Macroscopically, it is the temperature at which the
equilibrium crystal shape (ECS) loses its facets. That is, the ECS will
have (macroscopically\refto{Macroscopic}) flat [100] facets for $T < T_R$
and will be rounded for $T > T_R$.\refto{DifferentRougheningTemperatures}
On a microscopic level, $T_R$ is that temperature at which the step
free energy for steps (of all orientations)
across the [100] facet goes to zero.
The step free energy is defined as the difference between the free
energy per unit area for an interface with a step and the free
energy per unit area of an interface without the step.  (Note that it
depends on {\it both} the orientation of the interface and
the orientation of the step.\refto{InPractice})
When this step free energy goes to zero, the interface roughens since
there is no free energy barrier to the creation of steps which can
then proliferate without bound.

In the context of equilibrium crystal shapes, Rottman and
Wortis\refto{RottmanAndWortis, TilingModel, Wortis}
have shown that the addition of NNN AFM bonds
introduces two new transition temperatures which we will denote by $T_{CR}$
and $T_{ER}$.\refto{DifferentNotation}
At temperatures $T < T_{CR}$, the ECS is
a cube with macroscopically sharp edges and corners
(see \Fig{EquilibriumCrystalShape}).  Above the corner rounding transition
temperature, $T_{CR}$, the corners of the cube (or, equivalently, the edges
near the corners) become rounded but at least part
of the edge remains sharp.  As the temperature is increased further, the
rounding near the corners spreads out along the edges until, at
$T_{ER}$, the entire edge of the crystal is rounded.
$T_{ER}$ is thus referred to as the edge rounding transition
temperature.\refto{FirstCommunication}  On the microscopic level, $T_{CR}$
is defined as the temperature at which the step free energy, $f_{[111]}$,
for a step across a [111] interface goes to zero.  Analogously, $T_{ER}$ is
defined as the temperature at which the free energy, $f_{[110]}$, of a
step across a [110] interface goes to zero.\refto{MoreDetailsOnThis}

Let us now consider how these transitions affect dynamics.  The
relation of the roughening transition to dynamics has been known since such
a transition
was first proposed.  In fact, the interest in $T_R$ was sparked by studies
of crystal growth from the melt.\refto{Burton}  Below
$T_R$, where growing crystals
have smooth facets, the growth is quite slow.  This is because the
growth proceeds via the nucleation of a group of atoms (``islands'') on the
surface.  Such an island
is only stable once a critical nucleus size has been reached, and thus
a large free energy barrier must be surmounted.  Above $T_R$, when the
surface is already rough, no such barrier exists.

We propose that the corner rounding temperature $T_{CR}$ marks a similar
change in dynamical behavior:  Below $T_{CR}$, there will be a free energy
barrier per unit length to depin a step from the edge of a domain, thus
leading to logarithmically--slow coarsening dynamics.\refto{WhyBarrierAboveToo}
How do we justify this claim that the dynamical transition temperature
and the corner rounding transition temperature coincide?
The justification comes from the recognition that the free energy
barrier we calculated in Subsection~C is (up to a geometric
factor\refto{GeometricFactor}) simply the step free energy, $f_{[111]}$,
for a step across a [111] interface.\refto{Justification}
The result presented there [Eq.\ \(TCR)]
for the corner rounding temperature $T_{CR}$ is an exact expression
in the $J_1/J_2 \to \infty$ limit.  An alternative calculation
of $T_{CR}$ yielding precisely the same result has been given by
Shi and Wortis.\refto{TilingModel}

\chaptertitle III. \hskip 1em NUMERICAL SIMULATIONS OF COARSENING

We will shortly turn to numerical simulations of the coarsening process itself.
First, however, it is worthwhile to reexamine
our argument for logarithmic coarsening in more detail.  In particular, we
would like to explain why the argument based on shrinking cubic
droplets may be more general than it first appears to be.  The question to
be addressed is this:
Having identified one special configuration in which there are
large barriers, why do we have reason to believe that such barriers will be
present for the configurations which will occur
during the coarsening process?\refto{PottsExample}

There are two points to be made in answering this question.
First, an initial droplet which is not cubic will
tend to shrink to a point where it has flat
faces and sharp edges and then get stuck.
We have confirmed this numerically by studying the time to shrink
spherical droplets.
The time it takes to shrink such a sphere of diameter $D$ is in
agreement with the expectation that the barrier is approximately determined by
the largest cube contained inside the sphere, namely one with
edge length $L$ close to (or a little less than) $D/\sqrt{3}$.

The second point is that the argument applies to a more general
situation than merely the shrinking of isolated droplets.
[This greater generality is necessary because
isolated droplets (or, in the language of percolation theory, ``clusters''),
except of very small size, are a rare occurrence in the 3D
Ising model.  The reason is that the
density of both up and down sites ($p = 0.5$) is well above the percolation
threshold ($p_c \approx 0.312$),\refto{Percolation} so the system consists
primarily of two infinite clusters.]
For example, a ``handle--like'' structure of
size $L$ on one of the infinite clusters should also have an energy barrier
to shrinking which is proportional to $L$.
We believe the fundamental point is that below $T_{CR}$
there is a free energy barrier per unit length to depin a step from an edge.
This means that edges will tend to remain sharp and that the
free energy barriers to flip the spins along such an edge will
diverge with the length of the edge.  Since the length of edges in
the system should be proportional to the characteristic length scale $L(t)$,
we expect that the free energy barriers will diverge with $L(t)$
and the growth of $L(t)$ will be logarithmic.

On the basis of these arguments and
upon our numerical simulations of shrinking cubic droplets, we believe that
the case for logarithmic coarsening is fairly
strong.  Ultimately, however, lacking a rigorous proof, we
must turn to numerical simulations of the coarsening process in order to
further test our hypothesis.

\beginsection A. Growth of the characteristic length scale

We study coarsening following an instantaneous quench from infinite temperature
to a temperature $T < T_C$.
Such a quench is implemented by starting with a random spin configuration at
time $t = 0$, and then evolving the system at temperature $T$,
using Glauber single--spin--flip dynamics [Eq. \(Glauber)].
Further details of the Monte Carlo algorithm are given in Appendix~C.

The most important quantity to monitor during coarsening is the characteristic
length scale $L(t)$.  In the literature,
many ways of measuring $L(t)$ are discussed.\refto{Amar,OtherMeasures}
If scaling is obeyed, any physically reasonable measure of $L(t)$ should show
the same behavior. (A discussion of the scaling of the correlation function is
given in Appendix~D.)
We choose one of the most common and convenient measures of $L(t)$, which
is to take $L(t)$ to be
proportional to the inverse of the total perimeter of domain boundaries.
(As is further elucidated in Appendix~D, this corresponds to measuring
the slope of the correlation function $C(r,t)$ near $r = 0$.)
That is, we define
$$
	L(t) \equiv {-E^{\rm NN}_0\over{(E^{\rm NN} - E^{\rm NN}_0)}}
	\ ,
	\eqno(MeasureOfCharacteristicLengthScale)
$$
where $E^{\rm NN}$ is the energy associated with nearest--neighbor bonds
only and ${E^{\rm NN}_0 \equiv -3 N J_1}$ is this energy in the ground
state.\refto{GroundStateOK}
The normalization of $L(t)$ is such that, on average, $L(t) = 1$ for the random
initial configuration.

\Fig{coarsening_2d.ps} shows the growth of $L(t)$
during coarsening in two dimensions.\refto{Averaging}
We see that for $J_2 = 0$, the simulation
results obey the $t^{1/2}$ law quite well
over the entire time.  (There {\it are} some small, but statistically
significant,
deviations at early times.)  Adding the NNN bonds
changes the behavior in exactly the manner which we expect:  After a
short period of relaxation on the shortest length scales, the system finds
itself in a configuration in which it must flip corners to coarsen further.
Since the energy barrier to flip such a corner is $4 J_2$, the system is
stuck and coarsens little on time scales $t \ll e^{4 J_2/T}$.  However,
at a time of order $e^{4 J_2/T}$, we see a dramatic upturn
on this log--log plot.  This is because on time scales $t \gg e^{4 J_2/T}$,
the energy barriers can be crossed and we see $t^{1/2}$ behavior.  This
confirms our prediction that the 2D model is a class--2 system.\refto{Lai}

Now let's consider \Fig{coarsening_3d_loglog.ps} which
shows the growth of the characteristic length scale in
three dimensions.  As in two dimensions, there
is a short period of fast relaxation on very short length scales.
After that, there is a period in which the growth is roughly a
power law $t^{n_{\rm eff}}$ with the effective
exponent $n_{\rm eff}$ for the domain growth (given by the slope on this
log--log plot) ranging from $0$ (total freezing) for
$T / J_2 \to 0$, to $\sim0.35$ for $J_2 = 0$.  For
$T = 2, 3$ and $4J_2$, there is considerable downward curvature
in the data at late times, suggesting a crossover to logarithmic growth.
No such downward curvature is evident for $J_2 = 0$ or for $T = 8 J_2$
(which is above the corner rounding temperature $T_{CR}$) until finite-size
effects lead to a quite sharp flattening out once $L(t)$ is roughly a third
the system size.

Two brief comments are in order.
First, let's consider the behavior for $J_2 = T = 0$:  The
data appears to obey a power law quite well over a full two decades in time
and yet the power is found to be about $0.35$, much closer to
$1/3$ than to $1/2$.  This is a surprising result since it is gospel that the
nearest--neighbor Ising model obeys the Lifshitz--Allen--Cahn law.
This anomalous exponent is discussed in more detail elsewhere.\refto{Thesis}
Here, it will suffice to say
that similar behavior (an exponent of $\sim0.37$) has been seen by
Amar and Family,\refto{AmarAndFamily} in the only other study
(as far as we know) of coarsening in the 3D Ising
model at zero temperature.  Since some of their calculations were
performed on very large ($512^3$) systems, we can be confident that this
is not a finite size effect.  On the other hand, we cannot rule out the
possibility that this is simply a very long--lived transient.
This may indeed be the case, but there exists the intriguing possibility
that there is truly a different exponent (or a breakdown of scaling
altogether\refto{Thesis,AmarAndFamily}) at $T = 0$.
(The numerical evidence suggests that the
$t^{1/2}$ law will be obeyed at late enough times for any nonzero
temperature.\refto{Thesis})

A second comment concerns the interesting behavior seen for
$T = 0.75 J_2$, where there are clearly steps in the data.  These steps are
indicative of the discreteness of the energy barriers in our system.
On time scales $t \ll e^{4 J_2 / T}$, any spin flips which raise
the energy are unlikely to occur.  Then, on time scales $e^{4 J_2 / T} \ll t
\ll e^{8 J_2 / T}$, energy barriers of $4 J_2$ are crossed regularly, but
those of $8 J_2$ are not, and so on.
The origin of these steps is thus the same as that which leads to the
change in behavior in the 2D model at a time of order
$e^{4 J_2 / T}$, except that
here there is more than one barrier height present in the coarsening system.
For quenches to somewhat higher temperatures, the steps are closer together.
By a temperature of $T = 2 J_2$, they have been washed out.

Now let us look more closely at the data for moderate values of $T/J_2$.
One could conclude from these results that at long times there will be
power law behavior with the exponent itself a continuous function of
the ratio $T/J_2$.  However, even neglecting the considerable downward
curvature at late times, this scenerio seems rather unlikely to us (although
certainly not forbidden) since we know of no proposals
that a coarsening system will have a continuous set of exponents,
$n(T)$.  A more likely
explanation is that this is a transient behavior and that at longer times
there will either be upward curvature with a return to $t^{1/2}$
behavior (as occurs in the 2D model) or slow
downward curvature compatible with our prediction of logarithmic growth.
Apparent power law growth with temperature--dependent effective
exponents has commonly been seen in Monte Carlo simulations of systems
which freeze at zero temperature. It occurs both in systems which are known
to obey $t^{1/2}$ or $t^{1/3}$ growth laws\refto{Lai,EarlierPotts}
at long times and those believed to obey logarithmic growth laws.\refto{Random}

\beginsubsection 1. Evidence for logarithmic growth of $L(t)$

We now ask whether the data in \Fig{coarsening_3d_loglog.ps}
is more compatible with a return
to $t^{1/2}$ behavior or with a logarithm.  There are several reasons
why we prefer the latter explanation.  First, one can estimate
the largest barrier heights involved in the coarsening process out to the
times studied by noting that at $t = 10^6$ MC steps/spin and
$T = 3 J_2$, the time is three orders of magnitude
larger than the time taken for equivalent coarsening in a system with
$J_2 = 0$. From this, we can
get an estimate of the energy barriers $E$ which are being crossed by setting
this ratio of time scales equal to $e^{E/T}$; this gives
$E \approx 20 J_2$.  Activation over such large barriers suggests a process
involving the cooperative
flipping of several spins.  In addition, as long as the data for $J_2 \ne 0$
has a smaller slope than that for $J_2 = 0$, this difference in time scales
(and the energy barriers it implies) will continue to grow.

There are two caveats which must be noted.  First, the barrier of $20 J_2$ is
that associated with flipping an edge of length $4$, while
$L(t)$ is about three times this large.  We believe this discrepancy
is due to the fact that $L(t)$ gives only an average length scale in the
system, whereas the active processes will be determined by the shortest
length since these will be the fastest.
That there are a distribution of lengths in the system is apparent
when we look at the spin configurations during the coarsening.

The second caveat is that there is, of course, a larger energy scale than $J_2$
in our system, namely $J_1$.
Our simulations do not go out far enough in time to rule
out the possibility that at long times the system can avoid the growing
$J_2$ barriers by going over large, but $L$-independent, $J_1$ barriers.
We cannot envision such a scenario, but have not {\it proven}
its impossibility.  In order to test the sensitivity of our results
to $J_1$, we increased the ratio of $J_1/J_2$ and find that the results change
only a little (with the downward curvature becoming slightly more pronounced).
This suggests that
activation over $J_2$--- and not $J_1$---barriers is what is
important here.  Nonetheless, not being able to carry out our simulations
to times which are much greater than any single--spin $J_1$ barriers in our
problem is probably the greatest weakness in the conclusiveness of the
numerical results.  Ultimately, the best evidence that $J_1$ is not relevent
will provided in Section~IV, where we introduce the tiling model.  In that
model, the energy scale $J_1$ is eliminated altogether in a natural way by
incorporating it into a constraint on the dynamics.

Now to the second, and primary, reason for believing the growth to be
logarithmic:  the downward curvature apparent in the data
at late times for $T = 2, 3,$ and $4J_2$.  It suggests that, as the shortest
lengths in the system become large enough
for our arguments to apply, there is a crossover to slower
growth.  In order to test whether the growth is in fact becoming logarithmic,
we show
two--parameter fits to the form $L(t) = a \, \log(t/t_0)$ over the last two
to three decades in time.  There is evidence of some systematic disagreement
(with the Monte Carlo data having less curvature than the fits), but
in general the fits are quite good, and are far superior to any straight
line (\ie power law) fits.  (Why is there still some systematic
deviation from logarithmic behavior at these latest times?  In part this
may be evidence that we are still not yet quite in the late time, scaling
regime; but some of the deviations may be accounted for by the fact
that in this scaling regime we would really expect the length scale to satisfy
a differential equation of the form
$$
	{\ddt L} \sim {e^{-f_B L /T} \over{L}}
	\ .
	\eqno(TrueEquation)
$$
\Ie we must include the $1/L$ curvature force term.  This equation
only asymptotically gives the
form $L(t) = a \, \log(t/t_0)$.  Indeed, we find that fits over the same range
of times to the full form obtained from integrating Eq.\ \(TrueEquation)
are better than the fits shown, having virtually no
systematic deviation.\refto{ThirdParameter})

The final reason for believing the long time growth to be logarithmic
comes from observing the spin configurations which occur during a quench.
\Fig{comparison_of_configurations} compares a 2D slice through
a system with $J_2 = 0$ to that with $J_2 \ne 0$, at two times during the
coarsening.
In contrast to spin configurations for the unfrustrated ($J_2 = 0$)
model, the configurations for $J_2 \ne 0$ are clearly ``blocky'',
with a strong preference for flat faces aligned along the simple cubic
directions.  Thus the system seems to be getting stuck in
the sort of configurations where our argument for logarithmic
growth should hold.\refto{BlockyNotEnough}

\beginsubsection 2. Tests for finite--size effects

We will now briefly consider another possible source of the downward curvature
seen in \Fig{coarsening_3d_loglog.ps} for $T = 2, 3$, and $4 J_2$: finite--size
effects.  It is apparent from the data for $J_2 = 0$ and $T = 8 J_2$ that
the effect of the periodic boundary conditions is to introduce downward
curvature once $L(t)$ gets large.  However, this curvature is
qualitatively different than what we see for $T = 2, 3$, and $4 J_2$.
It is quite sharp rather than gradual, and, as a result, it
cannot be fit to a logarithmic form over any reasonable range of
times.\refto{Thesis}
Furthermore, it occurs at a considerably larger value of $L$.
[In particular, compare $J_2 = 0$ to
$T = 3 J_2$, where both runs are on $(110)^3$ systems.]
Nonetheless, the possibility that the curvature could be due to finite--size
effects is certainly worth investigating more closely.

Why do periodic boundary conditions produce a cutoff in the growth of $L(t)$?
This cutoff is due to the development of flat
slabs and tubes extending the entire length of the system (and thus,
in a sense, infinite), as has been discussed by others.\refto{Tubes}
These configurations are metastable
since there are barriers of at least $4 J_1$ to flipping any spins.
Because we are studying this model at temperatures $T \ll 4 J_1$,
these metastable states are very long--lived.
Furthermore, they are much more common in three dimensions than in two, since
the fraction of each type of spin, $p = 0.5$, is far above the percolation
threshhold for three dimensions but coincides with it for two dimensions.

We have studied finite--size effects in two ways.
First, we have tried changing the boundary conditions from periodic to
antiperiodic (in all three lattice directions).  This has the effect of forcing
an interface in the system and thus should rid us of the slab and tube
effects.  (Of course, such boundary conditions presumably introduce some
of their own idiosyncrasies---We don't claim these boundary conditions are
better, but only that they are different.  In particular, since an interface
is forced in the system, the largest possible value that $L$ [as defined by
Eq.\ \(MeasureOfCharacteristicLengthScale)] can assume in an ${\cal L}^3$
system is ${\cal L}/2$,\refto{CorrectGroundState}
so there will still be a cutoff in $L$ due to finite size effects.)
With these boundary conditions, there is no detectable change in the curvature
for $T = 3 J_2$.
Second, we have investigated the effect of varying the system size.  We have
done this most systematically for $T = 3 J_2$, where we find that there seems
to be no statistically--significant difference between a $110^3$
and an $80^3$ system, out to at least $t = 10^5$ MC steps/spin.
For a $55^3$ system (where the run-to-run variation at
late times is quite large and the results correspondingly less
trustworthy), the data has a tendency to have a bit {\it less}
curvature than for the runs at larger system sizes.

On the basis of these tests, we conclude that it is extremely unlikely that the
downward curvature in \Fig{coarsening_3d_loglog.ps} for $T = 2, 3$, and
$4 J_2$ is due to finite--size effects.

\beginsection B. Summary of simulation results

The main purpose of this Section has been to present the numerical evidence
in support of our
claim of logarithmic growth of the characteristic length scale $L(t)$ at
asymptotically long times.  Although evidence from numerical simulations
can never be {\it totally} conclusive, we believe that it is compelling.

There are, however, a few drawbacks to the numerical results.  One is that
it is difficult to image the complicated interpenetrating domain stuctures
in our 3D model.  Better visualization of the spin
configurations than can be obtained
from 2D slices would be helpful.  A second and
more important limitation is that our simulations do not go far enough out
in time to exclude the possibility that the growth of $L(t)$ will change
qualitatively (\ie speed up) once the time scales are such that
barriers of $8 J_1$ or larger can be crossed regularly.  Since such
barriers could correspond to single spin flips, we have not shown conclusively
that the free energy barriers to coarsening will continue to diverge
with length scale.

In the next Section, we study coarsening in a 2D
``tiling'' model closely related to our 3D model.
The tiling model does not suffer from either of the two drawbacks noted
above, and thus allows us to marshal even stronger support in favor of our
claim that systems without randomness can have logarithmically slow growth
of order.

\chaptertitle IV. \hskip 1em THE TILING MODEL

\beginsection A. Introduction to the tiling model --- statics

In this Section, we study the dynamics of what we dub the
``tiling model.''\refto{TilingModel,Blote,Veit}  This is a
2D model for a [111] interface in our 3D model.
A sample configuration of this model is shown in \Fig{sample_tiling.ps}.
The model views the interface from the [111] direction, which means that we are
seeing a corner between the [100], [010], and [001] facets head on.

In the statistical mechanics literature,
this model is known as the [111] restricted solid--on--solid
([111]-RSOS) model.
``Restricted'' here refers to the fact that in this model interface
configurations are restricted to those in which the entire interface is
visible when viewed from the [111] direction.  (\Ie configurations with
``overhangs'', an example of which is shown in \Fig{illegal_tiling.ps},
are forbidden).
It is the absence of overhangs which allows the interface to be represented
by a tiling of the plane with $60^\circ$ rhombi of three
different orientations.\refto{EqualTiles}
To make the energies of the tiling configurations correspond
to those of an interface in our 3D model with
AFM NNN bond strength $J_2$, we assign an energy of
$2 J_2$ for each unit length of boundary between unlike tiles.\refto{ItFollows}
Note that since the RSOS restriction forces
interfacial area to remain constant, the tiling model
corresponds to the limit $J_1 \to \infty$ in the 3D model.

At low temperatures, we expect the tiles to phase separate,
representing the fact that only a sharp corner in the 3D model is
thermodynamically stable.
At high temperatures, the different types of tiles should intermingle to form
a thermodynamically rough [111] interface.
Clearly the phase transition in the tiling model will occur at the point where
the sharp corner rounds, \ie at the corner--rounding transition $T_{CR}$ of
the 3D model.
(Although this is in fact the temperature for the order--disorder transition
in the tiling model, we will refer to it as $T_{CR}$, instead of $T_C$, in
order to remind the reader of it's connection to the 3D model.)
In fact, following the introduction of a related model
by Bl\"ote and Hilhorst,\refto{Blote}
the tiling model was used by Shi and Wortis\refto{TilingModel}
to study the corner rounding transition of a sodium chloride
(NaCl) crystal in equilibrium with its vapor.  Much of our introductory
discussion here follows closely that in Ref.\ \cite{TilingModel}.

An important feature to notice about this model (which is also true of other
restricted solid--on--solid models\refto{Jayaprakash})
is that excitations can only occur along the boundary between the
domains.\refto{Why}  Below $T_{CR}$, three ``frozen phases'' coexist,
corresponding to the microscopically flat [100], [010], and
[001] facets; the entropy per tile is zero in the
thermodynamic limit.  Of course, in any finite system,
there will be thermal excitations along the boundaries between domains.

We have explained how the tiling model is viewed either as a
2D tiling of the plane by rhombi or as the projection of an interface
in three dimensions.  Before discussing the dynamics of this model,
we will briefly discuss a third representation\refto{TilingModel,Blote} (which
is useful for implementing the dynamics):
an Ising spin system on a triangular lattice.  The mapping from a tiling
configuration to the corresponding spin configuration, examples of which
are shown in \Fig{tiling_dynamics}, is made by placing a spin at each vertex
such that the spins are antiferromagnetically
aligned along the edges of the rhombi.  The mapping is unique up to an overall
inversion of the spins.

Any spin configuration produced by a perfect tiling has, for each elementary
triangle, two out of three spins aligned.  Thus, the correspondence is
between tilings and ground state configurations of the NN
AFM Ising model on a triangular lattice.  The energetics
of the tiling and the corresponding spin configuration is reproduced if we
choose $J_2$ to be the bond strength for AFM
NNN bonds between the spins.
Thus, the tiling model with an energy cost of $2 J_2$ per unit length
of domain boundary can be represented by the following
Ising Hamiltonian on a triangular lattice:
$$
        H = J_1 {\sum_{\rm NN} s_i s_j} + J_2 {\sum_{\rm NNN} s_i s_j}
        \ ,
        \eqno(TilingHamiltonian)
$$
with $J_1 \to \infty$, $J_2 > 0$, and $s_i = -1$ or $+1$.
Since the spin configurations of \(TilingHamiltonian) are far less compelling
visually than the corresponding tilings, particularly in
making the connection with our 3D model, we will make only
occasional reference to this representation from here on.
However, it is, the representation which is computationally the most
convenient in performing the Monte Carlo simulations.

\beginsection B. Dynamics

We will use Glauber single--spin--flip dynamics [Eq. \(Glauber)]
on the spins of Eq.\ \(TilingHamiltonian).
Since we are in the $J_1 \to \infty$ limit,
the only spins that can flip are those which
have three of their six nearest--neighbors aligned.
In tiling language, such a spin flip corresponds to rotating an elementary
hexagon consisting of three tiles, as shown in \Fig{tiling_dynamics}.  Finally,
in the original 3D Ising description, it corresponds to
adding or removing an elementary cube (\ie to
flipping the spin that this cube represents).

{}From \Fig{sample_tiling.ps}, we see that there are relatively few
such elementary hexagons, and thus few sites
where this dynamical move can take place. This is, of course,
a consequence of the RSOS restriction (or, in terms of the corresponding
3D model, because we are in the limit $J_1 \to \infty$).

It is important to note that, even though this model represents an
interface of the 3D model, when we study coarsening in the two systems we
are looking at different processes.
In the 3D model, we studied coarsening for an
entire 3D system, not just a single interface.  In particular,
we were interested in
the decrease in the total interfacial area over time.  It was
the inverse of this total area which gave us our characteristic length scale.

Here, by contrast, we are confining ourselves to one particular
interface of {\it fixed} area.  Rather than studying how total
interfacial area shrinks, we are instead concerned with how the structure
of this one interface itself coarsens.  To that end,
we will quench the system from a temperature
where the [111] interface is thermodynamically stable in the rough phase to a
temperature where
it is unstable and thus reconstructs into pieces of [100], [010], and [001]
facets.\refto{WideLiterature}
We then look at the coarseness of this reconstructing interface over time.

In the 3D model, the order parameter was not conserved by
the dynamics, and the naively--expected power law was $L(t) \sim t^{1/2}$.
However, because the coarsening involved activation
over barriers that grow with $L(t)$, logarithmic growth was found to occur.
In the tiling model, the order parameter (measuring
the local tile orientation) is conserved by the dynamics, as can be seen from
\Fig{tiling_dynamics}, and thus the naively--expected power law is now
$L(t) \sim t^{1/3}$.
However, the method by which the interface coarsens involves activation
over precisely the same sort of energy barriers
as in the 3D model, and thus the same arguments we made
for logarithmically slow coarsening in that model will
apply here as well.\refto{CubicalProjections}

Although the tiling model is a bit more obscure than the Ising model
on a cubic lattice, it has several advantages over our 3D model.
We have already alluded to these advantages a few times previously,
but at the risk of boring the reader, we reiterate them, this time in more
detail:

\item{(1)}  Since the tiling model is two--dimensional, the configurations
are considerably simpler and easier to visualize.
Furthermore, since there are three types of tiles, each
has a density of $p = 1/3$, which is well below the percolation threshhold.
As a result, the domains are compact, with
structures much closer in form to cubical projections\refto{CubicalProjections}
than the convoluted domains in our 3D model are to cubes!

\item{(2)}  The simulations in 3D go out far enough
in time to show that the system should be crossing energy barriers which are
large in comparison to $12 J_2$, the largest $J_2$ energy cost
for flipping a single spin. However, the times are not long enough that
single--spin--flip barriers of $8$ or $12 J_1$ are being crossed.
Thus the simulations themselves do not exclude the possibility that,
once barriers of $8$ or $12 J_1$ could
be crossed, the system would get ``unstuck'' and coarsen like $t^{1/2}$.
Therefore, strictly speaking,
our numerical work really only proves the claim of logarithmic coarsening
in the limit of $J_1\to\infty$.  Since this limit corresponds to
$T_C\to\infty$, our numerical work is most definitive in the singular limit
$T/T_C\to 0$.  The tiling model eliminates the possibility of activation over
$J_1$-barriers by replacing the energy scale $J_1$ entirely by a
constraint on the dynamics.  Admittedly this constraint is equivalent to
the limit $J_1\to\infty$, but such a limit does not seem unreasonable since
the order--disorder transition, $T_{CR}$, in this model is proportional to
$J_2$, not $J_1$. Thus our argument for logarithmic coarsening can be
tested numerically over a large interval of $T/T_{CR}$. (We expect it
to hold all the way up to the transition!)

\item{(3)}  Finally, one might imagine that by eliminating one spatial
dimension,
we should be able to get even more convincing numerical results with the
available amount of computing power.   In fact, we will see that this does
not appear to be the case.  As with the 3D model, the slow
dynamics which we are studying also slows the approach to the
scaling regime where $L(t)$ assumes its asymptotic behavior.  Nonetheless, we
feel the numerical evidence is compelling enough that it would truly be
perverse if $L(t)$ resumes the naively--expected $t^{1/3}$ growth at
times beyond the longest times we reach in these simulations.

We are now ready to present the numerical results for the tiling model.
Since the presentation closely parallels that for the
3D model, we will be more terse here.

\beginsection C. Simulations of coarsening in the tiling model

We study the coarsening following a quench from
infinite temperature (\ie a random tiling) to a final temperature
$T$.\refto{RandomInitialTiling}
Once again, we take the characteristic length scale $L(t)$ to be
inversely proportional to the total perimeter of boundary between the domains.
In particular, we define
$$
        L(t) \equiv {-2.5 E_0\over{E - E_0}}
        \ ,
        \eqno(TilingMeasureOfCharacteristicLengthScale)
$$
where $E$ is the energy, and $E_0 \equiv -N J_2$ is the energy of the ground
state (\ie where the total perimeter of domain boundaries is
zero\refto{CorrectGroundState2}).  The arbitrary
constant $2.5$ is chosen for convenience so that
$L \approx 1.0$ for the initial random tiling.

\beginsubsection 1. Configurations during coarsening

Before studying the growth of $L(t)$ in quantitative detail, let us first
discuss the qualitative features of the configurations as they evolve
over time.  \Fig{tilings_during_coarsening} shows the evolution of the domain
structure following a quench to $T = 3 J_2$.  We see the expected sharp
boundaries between domains.  As the length of these sharp edges increases,
the energy barriers which must be surmounted to further coarsen the system
should increase.

One might argue that there has been significant
coarsening of the domain structure over the time period shown.  However,
it important to notice the large range of time scales over which these
snapshots have been taken.  If the system were coarsening like
$L(t) \sim t^{1/3}$, then the characteristic length scale would have grown
by a factor of almost 50 between $t = 10^4$ and $t = 10^9$ MC steps/spin.
Instead, it has grown by only a factor of 3.

In addition to studying still snapshots like these, we have
also made an animated ``movie'' of the evolution of the system.  Such a movie
shows compellingly the effect of the large activation barriers.  At late
times, the system
spends nearly all of its time ``climbing'' part of the way up these barriers
(\eg flipping a few spins along the edge of a domain) and then falling back
down to its original state.  Only rarely does it succeed in surmounting
the barriers and finding a lower energy state.

\beginsubsection 2. Growth of the characteristic length scale

Now we will examine the growth of $L(t)$ quantitatively.
\Fig{coarsening_tiling_loglog.ps} is a log--log plot showing the growth
of the characteristic length scale over time.  We see the same basic trend that
we saw for the 3D model ({\it c.f.} \Fig{coarsening_3d_loglog.ps}).
At the lowest temperatures, we see the steplike behavior in $L(t)$ which occurs
as the system reaches time scales when it is first able to surmount barriers
of height $4J_2$, then $8J_2$, and then $12J_2$.
For higher $T/J_2$, these
steps get washed out and the effective exponent $n_{\rm eff}$
(the slope on this plot), increases toward $1/3$.  For a given value of
$T/J_2$, $n_{\rm eff}$ appears to be slowly decreasing over time at late times.
(For $T = 4J_2$ and especially for $T = 5J_2$, $n_{\rm eff}$ is lower than
for $T = 3J_2$ at early
times, but then increases at moderate times before appearing to decrease again
at late times.  The behavior at early and moderate times is a result of the
sensitivity of our measurement of $L(t)$ to thermal fluctuations along the
domain boundaries. This will be discussed further in Subsection~3.)

The most sensitive test of whether the growth of $L(t)$ is becoming
logarithmic at late times,
is to study a plot of the slope, $dL/d[\log(t)]$, of the Monte
Carlo data when shown on a {\it log--normal} scale.  On such a plot,
an approach to logarithmic growth would be indicated by having
$dL/d[\log(t)]$ level off, \ie become constant, at late times.
Such a plot (see Ref.\ \cite{Thesis}, Fig.~4.12) shows that the
slope appears to be levelling off only a little for
$T = 2 J_2$ and $T = 3 J_2$.  For $T = 4J_2$, the slope is
levelling off more dramatically; however, since thermal fluctuations are
quite important at $T = 4 J_2$ even out to quite long times (as
discussed in Subsection~3), this result should be regarded with a bit
of skepticism.  We must conclude that out to the times studied in
\Fig{coarsening_tiling_loglog.ps}, we have
not yet reached the time regime when the growth is clearly logarithmic.

In order to better determine the long time behavior of $L(t)$, we have
carried out a few individual runs to times as long as $t = 2 \times10^9$ MC
steps/spin.  The results are presented in \Fig{coarsening_tiling_longtime.ps}.
Also shown are logarithmic and power law fits to the Monte Carlo data of
\Fig{coarsening_tiling_loglog.ps} over the final decade in time
($t = 10^6$--$10^7$ MC steps/spin for $T = 3 J_2$, and $t = 10^5$--$10^6$
MC steps/spin for $T = 4J_2$).
Since there are large run--to--run fluctuations, it is hard to come to any firm
conclusions.  However, the data generally seem to lie between
the logarithmic and power law extrapolations.  This suggests that although the
data still aren't quite fit by a logarithm, the exponent for the power law
fit (which is already very small) is continuing to decrease.
This gives us confidence that the growth at
asymptotically long times will in fact be logarithmic (or at least
slower than power law).  Unfortunately, the
same barriers which produce this slow growth appear, not surprisingly,
to slow the approach to this asymptotic behavior.

In order to show that the free energy barriers to coarsening at these late
times clearly involve the flipping of many spins,
we estimate a lower bound for these barriers as follows:
We consider the form $L(t) = a (t/\tau)^{1/3}$ with $\tau = \exp(F_B/T)$
and $a$ of order 1.\refto{InEffect}
Using the Monte Carlo data at the longest times gives
$F_B \approx 40J_2$ and $55J_2$, for $T = 3 J_2$ and $4 J_2$, respectively
(which is roughly the barrier heights we would expect given the values of $L$).
By comparison, the energy barrier to flip any single spin in this model
is {\it at most} $12 J_2$, and the barrier per unit length of edge is only
$4J_2$.

\beginsubsection 3. Effects of thermal fluctuations and finite system size

In Subsection~2, we noted that
for $T = 4 J_2$ and $5 J_2$ (and for $T = 3 J_2$ at very early times),
there is some upward curvature in
\Fig{coarsening_tiling_loglog.ps} at short and intermediate times.
The reason for this is that when domains sizes are small, the
thermal fluctuations along the domain boundaries lengthen the total perimeter
significantly and thus decrease the characteristic length scale $L(t)$ as
measured by the inverse of this total perimeter
[Eq.\ \(TilingMeasureOfCharacteristicLengthScale)].
This can be seen clearly in \Fig{tilings_thermal_fluctuations}.

We have tried to correct for these thermal fluctuations
by measuring $L(t)$ only after first quenching the
system to $T = 0$,  or by using different measures of $L(t)$ which we hoped
would be less sensitive to the thermal fluctuations,\refto{ExampleOfAttempt}
but none of these methods were completely effective.\refto{Thesis}
Since we have not yet found an adequate way to correct for
thermal fluctuation effects, we must ask to what extent they make
our measurement for $L(t)$ untrustworthy.
Fortunately, since the fluctuations occur only along the domain
boundaries, the error in the measured length scale should
decrease as the domain sizes grow.
{}From \Fig{tilings_during_coarsening}, our study of alternate definitions
of $L(t)$, and our studies of measuring
$L(t)$ after quenching to $T = 0$, we conclude that the data for $T = 3 J_2$
should be trustworthy at least for times $t > 10^5$ MC steps/spin.
For $T = 4 J_2$ and
$T = 5 J_2$, the situation is more unclear.  However, even the $T = 4 J_2$
results are probably quite trustworthy at the latest times.

Finally, a few words on finite--size effects.
To test for finite--size effects, we have run simulations for $T = 3 J_2$
on a system of half the linear size ($60^2$).  A comparison of the results
to those shown in \Fig{coarsening_tiling_loglog.ps} shows no evidence of any
systematic deviation.   The Monte Carlo data for the two systems agree within
the (quite small) error bars.  Since even the $60^2$ system shows no evidence
of finite--size effects out to at least $L \approx 7$, we imagine the results
on $120^2$ systems should be trustworthy out to at least
$L \approx 14$.  Thus, all the
results we have presented should not have any significant finite--size effects.

\beginsection D. Growth of order during slow cooling

In our 3D model, there are two distinct temperatures.
One is $T_C$, below which ordering occurs.  The other is $T_{CR}$, below
which free energy barriers are proportional to the length scale, thus slowing
the dynamics.
Since $T_{CR} < T_C$, this model orders without difficulty if it is
cooled slowly.  It must be {\it quenched} from $T > T_C$ to $T < T_{CR}$
in order to exhibit the logarithmically slow growth of order.

However, in the tiling model the ordering temperature and the temperature for
slow dynamics coincide.  Naively, we might hope that such a model will truly
behave like a glass in the sense of having difficulty ordering even under
slow cooling.  More precisely, in a glass the relaxation time scales might
diverge exponentially so that the final ($T = 0$) size of the correlated
regions would grow only like
$$
 	L(T=0) \sim \log(1/\Gamma)
	\ ,
	\eqno(LAnnealGlass)
$$
where $\Gamma$ is the cooling rate defined by the cooling schedule
$$
	T = T_{init} - \Gamma t
	\ .
	\eqno(CoolingRate)
$$

Does Eq.\ \(LAnnealGlass) hold  for the characteristic length scale in the
tiling model?
\Fig{annealing_tiling.ps} shows the growth of the characteristic length scale
$L$ upon cooling at a rate $\Gamma$ from an initial temperature $T_{init}$
somewhat above $T_{CR}$.  We see that the length scale
grows rapidly in a rather small region of temperature below $T_{CR}$.  (Recall
that since our method of measuring $L$ is sensitive to thermal fluctuations,
the true characteristic length scale is larger than that which is
measured at temperatures $T$ close to $T_{CR}$.)  Now let
us look at the final length scale reached as a function of $\Gamma$,
as shown in \Fig{final_annealing_tiling.ps}.  Since
\Fig{final_annealing_tiling.ps} is a log--log
plot, we expect downward curvature if the dependence of $L$ on
$1/\Gamma$ is logarithmic.  Clearly this is not the case here.  In fact, there
is some upward curvature at fast cooling rates with some evidence
that the data may be approaching a straight line (power law dependence) for
the slowest coolings.

{}From the point of view of modeling a glassy system,
these simulation results are disappointing, though not surprising
once we consider the cooling problem more closely.  Assuming that,
for times $t$ at which $T < T_{CR}$,
the characteristic length scale satisfies a differential equation of
the form
$$
        {\ddt L} \sim {e^{-f_B L /T} \over{L^2}}
        \eqno(TrueEquationTiling)
$$
[with $f_B$ given by Eq.\ \(FreeEnergyPerLength2)],
we will now derive an expression for the expected dependence of $L(T = 0)$ on
the cooling rate $\Gamma$ which should be valid
in the limit of asymptotically slow cooling rate ($\Gamma \to 0$).
This expression will show
that (even with the assumption that the barrier to domain growth is
proportional to the characteristic length scale) the final length scale on slow
cooling should have a power-law, not logarithmic, dependence on the cooling
rate.
The growing free energy barriers do, however,
change the exponent of the power law from that which we would have expected in
their absence.

To start, we note that
the characteristic length scale below $T_{CR}$ should initially grow like
$$
        L(t) \sim t^{1/3}
        \ ,
        \eqno(LAnneal)
$$
where
$$
        t = {1\over{\Gamma}} (T_{CR} - T)
	\eqno(TimeAnneal)
$$
is the time the system has been below $T = T_{CR}$.
However, once the barriers get large, the growth of $L$ will slow dramatically.
To a first approximation, we will imagine that there is a sharp cutoff.
That is, we assume $L$ grows like Eq.\ \(LAnneal) down to a certain
temperature $T_f$ and then freezes.
\Fig{annealing_tiling.ps} suggests this approximation is not too bad:  The
cutoff in the growth of $L(T)$ is quite sharp, particularly at slow cooling
rates.  Substituting \(TimeAnneal) with $T = T_f$ into \(LAnneal),
we find that
$$
        L(T_f,\Gamma) \sim \biggl [{1\over{\Gamma}} (T_{CR} - T_f)
	\biggr ]^{1/3}
        \ .
        \eqno(LAnneal2)
$$

What should we choose for $T_f$? One estimate is the temperature at which
the ratio of the barrier heights to the temperature is of order 1:
$$
        {L(T_f,\Gamma)f_B(T_f)\over{T_f}} = 1
        \ .
        \eqno(CutoffAnneal1)
$$
This is most certainly an overestimate of $T_f$, since we expect that $L$ will
still be growing when the barriers are this small.  A second estimate
is to set the time to surmount the barriers equal to the inverse of the
cooling rate:
$$
        e^{L(T_f,\Gamma)f_B(T_f)/T_f} = 1/\Gamma
        \ .
        \eqno(CutoffAnneal2)
$$
Certainly, we do not expect much further growth once the barriers have gotten
this large and in fact the growth in $L$ would have already slowed down
considerably by this temperature.  This is, therefore, probably an
underestimate of $T_f$.  We believe these two cutoffs should in fact
provide rigorous bounds on the asymptotic form of ${L(T = 0,\Gamma)}$.

For now we will use the latter estimate [Eq.\ \(CutoffAnneal2)],
although we will find our final
result to be very insensitive to the choice of the cutoff $T_f$.
To make further progress, we will also assume
$$
	{T_{CR}-T_f\over{T_{CR}}} \ll 1
	\ .
 	\eqno(TfAnneal)
$$
This assumption is justified once the cooling rate becomes small, as
can be seen from \Fig{annealing_tiling.ps} and will be verified
self--consistently at the end of this calculation.  We make use of
\(TfAnneal) to replace $T_f$ by $T_{CR}$ in the denominator of the exponential
in Eq.\ \(CutoffAnneal2).  We also use \(TfAnneal) to write
$$
	f_B(T_f) = a \> (T_{CR} - T_f)
	\ ,
	\eqno(fAnneal)
$$
expressing the fact that the free energy barrier per unit length grows
linearly with decreasing temperature near $T_{CR}$ (see
\Fig{free_energy_per_unit_length}).
Substituting \(fAnneal) into \(CutoffAnneal2)
allows us to solve for $T_{CR} - T_f$:
$$
	T_{CR} - T_f = {T_{CR}\over{a \> L(T_f,\Gamma)}} \> \log(1/\Gamma)
	\ .
	\eqno(TCTfAnneal)
$$
Finally, we substitute \(TCTfAnneal) into our expression for
$L(T_f,\Gamma)$, Eq.\ \(LAnneal2), and solve.
Making use of our assumption that $L(T = 0,\Gamma) = L(T = T_f,\Gamma)$,
we arrive at our final result:
$$
        L(T=0,\Gamma) \sim \Biggl [{\Gamma\over{\log(1/\Gamma)}}\Biggr ]^{-1/4}
        \ .
        \eqno(LAnneal3)
$$

If we had used \(CutoffAnneal1) instead of \(CutoffAnneal2), we
would have obtained the same result without the factor of $\log(1/\Gamma)$.
This shows that the result is very insensitive to our exact choice of
cutoff and, in particular, that this choice should not change the exponent.
Thus, we expect that in the limit $\Gamma \to 0$, and
up to corrections logarithmic in $\Gamma$, the
characteristic length scale at $T = 0$ is\refto{AsymptoticallyCorrect}
$$
        L(T=0,\Gamma) \sim \Gamma^{-r}
        \ .
        \eqno(LAnneal4)
$$
with
$$
	r = 1/4
	\ .
	\eqno(ExponentAnneal)
$$

Note that if $T_f$ were roughly independent of $\Gamma$ (as would be
true if the barrier heights did not depend upon $L$),
we would have obtained simply $r = 1/3$,
since the time $t$ that $L$ has to grow would go like $1/\Gamma$.
The change to $r = 1/4$ is a nontrivial result of the growing barriers.
It reflects the fact that $T_f$ rises toward $T_{CR}$
as the cooling rate becomes smaller, because
the larger $L$ means larger barriers at a given temperature.\refto{OtherPowers}
In fact, Eq.\ \(TCTfAnneal) and \(LAnneal3) yield
$$
        T_{CR} - T_f \sim \Gamma^{1/4}
        \ ,
        \eqno(TCTfAnneal2)
$$
which confirms the validity of assumption \(TfAnneal) in
the limit $\Gamma \to 0$.

On the basis of this calculation, we conjecture that, in general,
a free energy barrier which goes {\it continuously} to zero at the
ordering transition should lead
to power--law, and not logarithmic, dependence of the zero--temperature length
scale on the cooling rate.  This
leads us to the conclusion that we need one of two things
in order to get a model with length--scale--dependent barriers which behaves
like our conception of a glass upon cooling:
$(i)$ A free energy barrier per unit length which jumps
discontinuously to a nonzero value at the ordering transition, rather than
rising continuously from zero; or
$(ii)$ a free energy barrier per unit length which is nonzero even above the
ordering transition.
The latter is what is believed to occur in the random field Ising
model.\refto{GrinsteinFernandez,FisherRFIM}
Whether nonrandom models with one or the other of these characteristics
exist remains an open question.

\chaptertitle V. \hskip 1em SUMMARY, CONCLUSIONS, AND OPEN QUESTIONS

The main goal of this paper has been to address the following question:
Can we have free energy barriers to ordering
which increase with the size of the correlated regions
in a system which does not contain disorder (\ie randomness in the
Hamiltonian)?   We have focussed our attention on
the growth of order in systems far out of equilibrium, in particular, systems
which have been quenched from $T=\infty$ (a random state) to a temperature,
$T < T_C$, at which the equilibrium phase has long--range order.

Expanding on the work of Lai, Mazenko, and Valls,\refto{Lai} who made
the underappreciated
point that free energy barriers proportional to the characteristic length scale
will lead to logarithmically--slow coarsening,  we have given two closely
related examples of models, free of randomness in their Hamiltonians,
in which such growing free energy barriers and logarithmic coarsening do,
indeed, occur.
Although we have not yet been able to rigorously {\it prove} that the
domain growth in the two models is logarithmic, our Monte Carlo results
strongly support our heuristic arguments that this is so.

The growing barriers arise because the mechanism by which these systems coarsen
involves the creation of a step across a flat interface.  Below
the corner rounding transition temperature, $T_{CR}$,
the creation of such a step costs a free energy proportional to its
length.  Thus, the coarsening dynamics in our 3D model
are logarithmically slow only below $T_{CR}$.  In the corresponding equilibrium
system, $T_{CR}$ marks an interfacial phase transition involving the rounding
of corners on the equilibrium crystal shape.
It has long been understood that a connection exists
between the roughening transition and the dynamics of crystal growth.
The connection between coarsening dynamics and the corner rounding
transition provides a natural extension of the relation between
interfacial phase transitions and growth dynamics.

Along the way, we have also studied the 2D Ising model with AFM NNN
bonds.  This is a particularly clean realization of a system with
length--scale--independent activation barriers, and thus serves as a canonical
example of such a ``class-2'' system.\refto{Lai}  Finally, we have also
discovered
(or actually rediscovered\refto{AmarAndFamily}) that even the nearest--neighbor
Ising model (no NNN bonds!) in three dimensions shows anomalously slow
coarsening at zero temperature.  This could be just a very long initial
transient, but there exists
the possibility that the exponent for the growth law is truly smaller
(or that scaling breaks down altogether) at $T = 0$.---This is
our one simulation result which still lacks a theoretical understanding.

\beginsection A. Frustration and the slow growth of order

We believe frustration to be an important feature in
producing the slow dynamics.  Here we would like to elaborate
in what sense we mean ``frustration.''  The
Ising model with NNN bonds is clearly frustrated in the
commonly--used sense that not all the interactions can simultaneously be
satisfied.  On the
other hand, this frustration does not manifest itself in complicated
equilibrium behavior: for
$J_1/J_2 > 2d$ (where $d$ is the dimensionality), the ground state is simply
ferromagnetic.

In what way then is frustration an important component of our models?
The sense of frustration which we are looking for
is more of a dynamic one:  For the models to
lower their free energy they must evolve through states of higher free energy.
That is, in order to increase their order on long length scales,
the models must first decrease their order on shorter length scales.
In the models we have studied here, this ``dynamic frustration'' comes about
because these models prefer sharp domain edges (\ie no steps), and in order to
coarsen, the system must necesarily pass through states in which the domain
edges are not sharp.  However, it seems likely that one can come up with
other ways (still without introducing imposed disorder) to dynamically
frustrate a system.

\beginsection B. Applications to experimental systems

There are two questions which arise in considering the applications of these
ideas to experimental systems.  Looking most narrowly, we can ask whether
there exist systems in which
the interactions might lead to slow coarsening of domains.
Taking a broader view, we can ponder what our work says more
generally about when a system might have diverging barriers associated with a
diverging length scale.
Since the latter question is what originally got us started on this work,
we will consider it first.

\beginsubsection 1. Relation to the glass transition

The original motivation for this work was to demonstrate the
plausibility of our speculations on the
glass transition (discussed in Appendix~A and, in more detail, in
Ref.\ \cite{OurGlassPaper}),
by searching for models without randomness which
nevertheless have free energy barriers that diverge with the
length scale over which the system is ordered.
We believe we have found such models.  Furthermore, we have also demonstrated
the more specific point that a low--temperature ordered phase itself can
have slow dynamics.  This result is important to explain why the
dynamics of a liquid cooled below the glass transition might remain slow even
away from the postulated second--order phase transition at $T_0$, where the
{\it equilibrium} correlation length should decrease back down to microscopic
values.  One might naively expect that this small correlation length means that
the ordering dynamics should speed up again, but we have demonstrated a
situation in which the relevant
length scale with which the barriers diverge is not the
equilibrium correlation length, but rather the length scale on which the
nonequilibrium system has ordered.  Thus, if a system
finds itself far out of equilibrium in the low--temperature phase,
its dynamics can remain slow even at temperatures well below the transition.

However, in one very important respect
the models which we have studied here do not themselves behave as our
conception of glasses: they do not have a logarithmic
dependence of the final ordering length on the cooling rate.
That is, the ordering is not as slow as we would expect when the system
is cooled slowly through the ordering transition.
The reason, as discussed in Subsection~IV.D, is that the free energy barrier
per unit length for the tiling model
goes continuously to zero as the ordering temperature is approached from below
(and for the 3D model, the barrier goes to zero at $T_{CR}$,
a temperature well below the ordering transition).
It seems likely from the analysis presented there
that any system for which this is true
will not show a logarithmic dependence of the length scale on cooling rate.

What do we need in a model to get a real glass?  The free energy barriers
in the models of this paper arise purely from {\it step} free energies.
As such, they are expected to go to zero at the transition temperature, as
in the tiling model, or even at a lower temperature, as in the 3D model.
Our conception of glasses is somewhat different.
We imagine that the growing order in a supercooled liquid
might consist, for example, of
small icosahedrally--ordered regions (``balls'').
The center of these balls would be
most tightly ordered.  Further out from the center of each
ball, the atoms would be more loosely held because of the
frustration induced by trying to
fill ordinary space with icosahedral order.   Note that, unlike in the models
considered in this paper, here the energy density in the
ordered regions is not uniform.
This nonuniformity is vitally important, since
it means that domain walls would like to sit preferentially in gaps
between the tightly ordered regions.  We believe that
moving a domain wall through a tightly ordered region could
then cost a free energy proportional to the size of the ordered region even
above the ordering transition.

This picture is closely analogous to the random field Ising model,
in which the free energy barriers are believed to diverge with the
diverging correlation length even above
$T_C$.\refto{FisherRFIM,ToOurGlassPaper} In this model, the nonuniformity is
the result of spatial fluctuations in the (quenched) random field variable.
Such fluctuations allow the free energy barriers per unit size to flipping a
domain to retain a nonzero value even above the transition temperature.  That
similar behavior can occur in a system without disorder remains to be shown.

\beginsubsection 2.  Relation to coarsening in experimental systems

Having discussed the broader question of free energy barriers which diverge
with any ordering length scale, let us now
consider the specific case we have studied, namely, barriers
which diverge with the characteristic length scale in a coarsening system.

First, we may ask the general question: Has unexplained slow coarsening been
seen before in {\it any} experimental systems?
It is usually found that for grain growth in
annealed metals and ceramics, the grain size $L$
can be fit reasonably well to the form $L(t) \sim t^n$.\refto{GrainGrowth}
However, the exponent
$n$ has often been found to be somewhat (and occasionally, considerably) less
than $1/2$, particularly at lower temperatures.  On the other hand, it must be
remembered that experimental systems can have slow coarsening at
low temperatures (and early enough times) simply
because, unlike in the Ising model, there are always
length--scale--independent
barriers involved with the elementary dynamical process of moving
atoms around.  Furthermore,
there has been no general trend for the exponent
to decrease at late times (at least, no such trend which cannot be explained
trivially by effects such
as the finite size of the sample).  Also, the general trend has been for the
exponent to be lowest in dirty materials and closest to $1/2$ for those of
high purity.  This suggests that at least many of the observed
low exponents are due to impurities.

At this point we should ask whether we
have any {\it specific} reason to expect such grain growth to be logarithmic.
In fact we do not, since
grain growth in metals and ceramics differs in several ways
from coarsening in the models which we have
studied.  Firstly, we have no reason to suspect that the interactions in
these materials are similar to those in our models. Secondly, in the grain
growth problem there is no superimposed lattice structure
like we have in our models.  Instead, the local orientation of the
lattice is determined by the local value of the
order parameter within the grain itself.

Rather than just diving headlong into the experimental literature in search of
slow growth, let us, instead, look for systems in which the
interactions are similar to those of the models we have studied, that is, in
which interactions produce domains with sharp edges and corners.
Equilibrium crystal shapes have been measured for only a few
materials.\refto{ExperimentalDifficulties}
Most of the materials studied do not exhibit
nonzero edge and corner rounding transitions.---The edges between
facets are rounded at any nonzero temperature (as they are in an
Ising model in which the NNN bonds are
ferromagnetic or zero).  The one exception is the
ionic salt sodium chloride, NaCl, whose crystal shape is strictly cubical
at low temperatures.  In fact, the work of Shi and Wortis\refto{TilingModel}
on the interface model of Section~IV was motivated by experimental
observations\refto{Heyraud}
of a corner--rounding transition at $T \approx 650^\circ C$
for NaCl crystals believed to be equilibrium with their
vapor.  Furthermore, some earlier experiments on the NaCl crystals actually
looked at
the coarsening of the $[111]$ interface of a NaCl crystal for temperatures
in the vicinity of $650^\circ C$.\refto{Knoppik}  Pictures of the
coarsening interface look strikingly similar to those we obtain for the
tiling model.\refto{LargeBody}

Is the slow coarsening dynamics we predict actually seen in these
experiments on NaCl?  The answer is unclear.
The coarsening length scale observed
after annealing does peak near the corner rounding transition and then drop at
lower temperatures.  These experimental observations prompted
Shi and Wortis to note that ``the fact that coarsening goes away as $T$ is
further lowered beyond [the corner rounding transition temperature] is
probably a kinetic effect (slow equilibration at lower temperatures).''
This sounds very promising but several strong caveats are in order.
The first is again the reminder that in this experimental system
there are (coarsening--length--scale--independent) barriers to the
elementary process of mass transport which might be large enough to cause slow
dynamics over the time scales of the experiment.\refto{Wortis}
The second caveat is that the time and temperature dependences of the
coarsening which are actually reported in the experiments are difficult to
interpret.
The final caveat is that the range and quality of the experimental results
are limited:  The experiments were only carried out in a rather narrow
range of temperatures within about $25\%$ of the corner--rounding transition.
Furthermore,
there exists the possibility that surface contamination or other factors
played an important role in what was seen.  (The experimentalists who studied
the equilibrium crystal shape itself\refto{Heyraud} note
that the degree of coarsening seen in these earlier experiments\refto{Knoppik}
is so large as to be inconsistent with their results.)

In the final analysis, we must conclude that while the experiments provide
some tantalizing
hints that interesting dynamics may be present, we cannot say that
they provide either support or refutation for our claim of
logarithmically slow coarsening.  It seems unlikely that the detailed
study of the dynamics necessary to detect the predicted logarithmically slow
coarsening will be made unless the
experimentalists know what they are looking for.  We hope
that future experiments on NaCl, or other crystals which have a corner
rounding transition, might look more closely at the dynamics of coarsening
of a [111] interface, and
particularly at the time--dependence of the coarsening length for temperatures
not too near the corner--rounding transition.
Such experiments will likely prove difficult to perform and it may be hard
to separate the effects discussed in this paper from those due to the large
elementary energy barriers associated with mass transport.  However, we think
the rewards of such an experiment make it worthy of
the attempt.

\beginsection C. Open Questions

Finally, we would like to leave the reader with some open questions
to ponder along two different lines:

\item{$\bullet$} How ubiquitous are systems with logarithmically slow
coarsening dynamics or, more generally, with free energy barriers to ordering
which diverge with the ordering length scale?  Have we found a few systems
which are exceptions or are such systems quite common?  What features are
important in producing such barriers?  In what experimental systems might we
expect to see slow dynamics?

\item{$\bullet$} Can we find models (without randomness) in which the free
energy barriers occur in such a way that they lead to slow ordering dynamics
even when a system is cooled slowly?  That is, can we find models which
truly fit our conception of glasses?

\acknowledgements

We are grateful to Veit Elser for suggesting that we study
the tiling model discussed in Section~IV.
We also thank David Huse, Peter Nightingale,
Teresa Cast\'an, Jennifer Hodgdon, Jacques Amar, and
David DiVincenzo for helpful discussions.  Finally,
JPS would like to thank Daniel Fisher for explaining his scaling
ideas about the glass transition,
which we've elaborated upon and published elsewhere,\refto{OurGlassPaper}
and which motivated the work presented here.  This work was supported in
part by NSF grant Nos. DMR 88-15685 and DMR 91-18065.
Computing facilities were provided in part by the Cornell--IBM Joint Study
on Computing for Scientific Research and the Cornell Materials Science Center.

\appendix A: \hskip 1em  SLOW DYNAMICS IN GLASSES

In this Appendix, we will briefly discuss our speculations about the glass
transition and how these speculations motivated us to search for models which
coarsen only logarithmically in time.\refto{ToOurGlassPaper}

When a liquid is supercooled below its freezing point, it gradually becomes
more and more viscous.  This continues until some temperature $T_g$
at which the time scales for
relaxation become so slow that the liquid can no longer equilibrate on
the time scales of the experiment.\refto{PseudoEquilibrium}
For all intents and purposes, the
system behaves as a solid, and yet it has no long range order.
It has been aptly described
as ``a liquid suspended in time.''\refto{FrozenInTime}
We call such a state a glass, and $T_g$ the glass transition temperature.

There has been little discernable progress in understanding the existence
of the glass transition despite over half a century of work on the problem.
Many theories have been presented but no real consensus seems to be forming.
There is one point on which everyone agrees---the ``transition'' at $T_g$
itself is purely a dynamical phenomenon.  In fact, the location of the
transition varies with how one chooses to define it.  A patient experimentalist
who is willing to wait longer to allow the system to relax (\ie who makes
a measurement on a longer time scale) would report a slightly lower value
for $T_g$.  For the purposes of uniformity, the location of $T_g$ is
conventionally
defined as that temperature where the viscosity reaches $10^{13}$ poise.

But {\it why} does the liquid become so sluggish?  One reason systems become
sluggish at low temperatures is because movement of the atoms involves
activation over free energy barriers.  In order to test for such activated
behavior, one can plot the logarithm of the viscosity $\eta$
(or any other measure of the relaxation time scale, such as the inverse of
the characteristic frequency of dielectric relaxation or
of the frequency--dependent specific heat\refto{Dixon})
vs. $1/T$.\refto{OurGlassPaper}  On such an Arrhenius plot,
activation over a constant barrier should produce a
straight line.  Indeed, for some materials, like SiO$_2$, the
data can be fit well with a straight line.  However, for most other
systems there is
clear upward curvature on the plot.  In fact, it has long been known that
viscosity data for glasses can often be fit reasonably
well\refto{HowWellFit} with the empirical Vogel--Fulcher Law:
$$
	\eta = \; \eta_0 \> e^{A/(T-T_0)}
	\ ,
	\eqno(VFLaw)
$$
where $T_0$ is typically tens of degrees below $T_g$. (For SiO$_2$, of course,
$T_0 = 0$ works quite well.)

What physics underlies this empirical law?
If we still want to interpret the data within the framework of free energy
barriers to relaxation (which is not the only possible
interpretation), we are forced to conclude that the barriers themselves
must be increasing as we lower the temperature.  If we believe Eq.\ \(VFLaw)
holds all the way down to $T_0$ then the barriers are in fact diverging.
The Arrhenius law yields Eq.\ \(VFLaw) if the barriers vary
with temperature as
$$
	E_{\rm VF}(T) = A \left|{T-T_0 \over T}\right|^{-\theta}
	\ ,
	\eqno(VFBarriers)
$$
where $\theta = 1$ gives the Vogel--Fulcher law exactly, although other
powers in the neighborhood of $1$ seem to fit data about equally well.

The power law divergence at $T_0$ in Eq.\ \(VFBarriers) should immediately
remind us of the divergences which would occur in many
measured quantities if $T_0$ marked a second order phase transition, and yet
the free energy barriers to relaxation is not usually one of those quantities
which diverges
at the transition.  In fact, in ordinary second order phase transitions,
it is the time scales to relaxation which diverge with a power law (so--called
``critical slowing down''); whereas Eq.\ \(VFLaw) implies that approaching
$T_0$, time scales are diverging exponentially fast!

Why might barriers be diverging at this transition?  Since the
correlation length is one of the traditionally divergent quantities at a
transition, it seems natural to assume that the barriers may be
growing with the size of the correlated regions.  In other words, perhaps
the supercooled liquid is trying to organize itself into some ordered state,
but gets stuck very quickly because of the growth of the barriers.

If such a scenario occurs in nature, why hasn't it been seen before?  The
answer to this question is twofold:  Firstly, the very existence of such
diverging barriers makes a close
approach to the equilibrium transition at $T_0$ impossible since the time
scales diverge exponentially fast!  The system will necessarily fall out of
equilibrium well before reaching $T_0$ (no matter how patient the
experimentalist is).  Thus, it is not surprising that these transitions
would be amongst the last to be understood.  Secondly, there is in fact at
least one system in which this scenario {\it is} believed to occur, namely the
3D random field Ising model and its experimental realization, diluted
antiferromagnets.  The dynamics in these systems is so slow
that one can never find the ordered state by cooling and
it took a rigorous mathematical proof\refto{Imbrie} to convince skeptics
that there is an equilibrium ordered phase at low
temperatures.\refto{Grinstein} It has been suggested,\refto{GrinsteinFernandez,
FisherRFIM}
and is now generally agreed,\refto{Lai} that the reason the dynamics is so
slow is that there are energy barriers to equilibration which diverge
with the size of the ordered regions.  Similar, but even more complex,
mechanisms may be at work in spin--glasses.\refto{Lai,FisherAndHuse}

However, there is a problem with blissfully applying these ideas
to glasses:  An important component in all the above--mentioned
systems is imposed disorder (\ie randomness in the Hamiltonian),
whereas in glasses, imposed disorder (due to
impurities, defects, or the like) is not thought to play a vital role.
But is imposed disorder really a necessary condition to get such diverging
barriers, or will a weaker condition suffice?

The motivation for the work presented in this paper was to investigate
the possibility that frustration without disorder can, at least in some
circumstances, produce diverging free energy barriers to relaxation and the
resulting slow dynamics.\refto{WhatFrustration}
We present heuristic arguments and strong numerical evidence for a
class of models that have barriers which diverge with the relevant
length scale associated with the size of the ordered regions.  However, this
length scale is not the equilibrium correlation length, but rather the
characteristic length scale in a coarsening system, that is, in a system
which has been quenched well below its transition temperature.  When cooled
slowly, our models are not glassy (See Subsection~IV.D).\refto{Shumway}
Nonetheless, the
discovery of such slow dynamics for nonrandom coarsening systems
is exciting in itself, and is also a large step toward
our more ambitious goal of producing models which behave in a glassy manner
even when cooled slowly.
Finally, it is worth noting that the work presented here, while motivated by
our speculative ideas about the glass transition, is in no way dependent upon
the correctness of these speculations.

\appendix B: \hskip 1em CALCULATING THE ENERGY OF CONFIGURATIONS

In this Appendix, we briefly describe how one goes about calculating the
energy of interface configurations for the 3D Ising model with
both nearest--neighbor (NN), $J_1$, and next--nearest--neighbor (NNN), $J_2$,
bonds. Recall that our convention is that $J_1 > 0$ and $J_2 > 0$
when the NN bonds are ferromagnetic and the NNN bonds are antiferromagnetic.
We compute the energy relative to a system with no interface.

Consider a unit area of (microscopic) interface between domains, a
``plaquette," as shown by the shaded square in \Fig{examples}.  Each such
plaquette is associated with a broken $J_1$ bond (energy cost of $2 J_1$)
which passes through its center.  Every plaquette also
has two $J_2$ bonds passing through each of its 4 edges.  Since each of these
edges is shared with another plaquette, we associate only four
of these eight $J_2$ bonds with each plaquette.  If the plaquette is part of
a flat interface then these 4 antiferromagnetic bonds (which are broken in the
bulk) will now be satisfied.  We therefore
associate with each plaquette an energy cost of $E_p = 2J_1-8J_2$.

Note, however, that there is an additional energy associated with each
(unit length of) bend in the interface (see \Fig{examples}).  The reason is
that along the edge of the plaquette where
the bend occurs, only one rather than both the $J_2$ bonds passing through
that edge will be satisfied.  This means the energy per unit length of
bend is $E_b = 2 J_2$.

We now summarize the two rules of energy accounting:

{\bf \item{1.} Each plaquette (unit area of interface) costs an energy
$E_p = 2 J_1 - 8 J_2$.}

{\bf \item{2.} A bend in the interface costs an energy per unit length of
$E_b = 2 J_2$.}

These two rules provide us with a simple
understanding of the $T = 0$ phase diagram of this model:
For $J_1/J_2 < 4$, the ferromagnetic ground state becomes unstable because the
energy per plaquette is now negative. This means that it becomes energetically
favorable to form interfaces between regions of up and down spins.  However,
it is still unfavorable to have bends in the interface, which is why the ground
state is striped, with alternating planes of up and down spins (and thus
flat, planar interfaces).

Also, we can see why the $J_2$ bonds lead to nonzero edge and corner rounding
transition temperatures.  The bend energy $E_b = 2 J_2$ produces an
attraction between steps, which stabilizes sharp edges and corners up to
nonzero temperatures.

\appendix C: \hskip 1em  IMPLEMENTATION OF THE MONTE CARLO ALGORITHM

The standard method of implementing Monte Carlo (MC) dynamics with random
updating proceeds as follows: One first chooses a spin at random, and then
flips this spin with an ``acceptance probability'' given by Eq. \(Glauber).
The traditional time unit is one MC step/spin, defined as $N$ such attempts,
where $N$ is the number of spins.

This method works very well for temperatures $T$ which are not too small
compared to the energy costs $\Delta E$ to flip most of the spins, so
that the ``acceptance rate'' (the fraction of the spin flip attempts which are
accepted) is reasonably large.  This
is definitely not the case in our system.  Rather, we are in the limit where
almost all the attempted flips will be rejected because most of the spins
have large energy costs $\Delta E \gg T$ to flipping.  For such situations,
a much more efficient algorithm has been proposed by
Bortz, Kalos, and Lebowitz,\refto{Bortz} the so--called ``continuous time''
Monte Carlo method.  It has been used previously to study
coarsening in Potts models.\refto{NoLogPotts}
The basic insight leading to continuous time MC
is that the standard MC method is hampered
by its fixed time step.  When few energetically--favorable flips are
possible, standard MC must adjust by lowering its acceptance
rate, whereas what one would like to do is to keep the
rate of acceptance high and, instead, compensate by incrementing the time
step by a larger amount.  By doing this correctly, one can actually
make every attempt a successful flip.   The method proceeds as follows:
\item{\bf 1.} Add together the acceptance probabilities for all the spins
in order to obtain a total flip rate $\Gamma$.  Increment the time by an amount
$\tau$ (in MC steps/spin)
where $\tau$ is a random number chosen from an exponential distribution
with mean $1/\Gamma$.
\item{\bf 2.} Choose a spin, with the probability for a given spin to be
chosen equal to its fractional contribution to the total rate.  Thus
spin $i$ with acceptance probability $\Gamma_i$ will be chosen with
probability $\Gamma_i/\Gamma$.
\item{\bf 3.} Flip this spin and go back to step 1.

The drawback of this method is that
each such flip entails quite a bit of overhead, particularly in the step~2.
In actual practice, as Bortz {\it et al.} pointed out, the algorithm can be
speeded up by sorting the spins into classes.  The idea is that at the start
of a run, one classifies each spin according to the number of
nearest--neighbors and next--nearest--neighbors aligned with it. (For the
3D model, there are 91 such classes.)  One then keeps a
table of the acceptance probability for a spin in each class.  The total rate
$\Gamma_j$ for a class $j$ is then this acceptance probability multiplied by
the number of spins in the class.
Choosing which spin to flip is now done in two stages.
First, a class $j$ is chosen with probability $\Gamma_j/\Gamma$.  Then
which spin in this class to flip is chosen totally at random (since all have
an equal acceptance probability).  One then reclassifies this spin and its
nearest-- and next--nearest--neighbors, calculates the resulting change
in the total rate $\Gamma$, and repeats the procedure of incrementing the
time and choosing the next spin to flip.

When the acceptance probabilities are high, this method still has enough
additional overhead to make it slower than the standard method.  Therefore,
for coarsening simulations we have often employed a hybrid of the two
algorithms:  At short times when the domains are small and the acceptance
rate for the standard MC method is high, we use the standard method.
At later times, we switch over to the continuous time method.
At the latest times, the savings over the standard method can be {\it very}
substantial.  For example, in a system of $40000$ spins, there may be
an average of less than one successful flip per time unit (\ie the
acceptance rate for the standard method is less than $1$ in $10^{4}$)!

The two methods yield equivalent results.  This can be seen by noting that
for both methods, the ratio of the occurrence of flipping two spins
is given by the ratio of their acceptance probabilities.  Also, for
both methods, the total number of flips per time step will, on average,
be given by $\Gamma$, the sum of
all the acceptance probabilities.  At early times, we
have compared the two methods numerically and find that they do indeed give
equivalent results within statistical errors.\refto{FoundError}

\appendix D: \hskip 1em SCALING AND ANISOTROPY OF THE CORRELATION FUNCTION

Here, we present a short discussion of Monte Carlo results for the
scaling properties of the correlation function, $C({\bf r},t)$, for the
model discussed in Sections~I---III.

The scaling hypothesis states that at late times
the correlation function $C(\bf r,t)$ is not a function
of $\bf r$ and $t$ separately, but rather
depends only on the ratio ${\bf r} / L(t)$:
$$
        C({\bf r}, t) = \tilde{C}({\bf r} / L(t))
        \ .
        \eqno(ScalingCorrelationFunctionAgain)
$$

The term ``scaling'' is also used to refer to the stronger statement
one makes by replacing $L(t)$ with its asymptotic form $t^n$, \ie
$$
        C({\bf r}, t) = \tilde{C}({\bf r} / t^n)
        \ .
        \eqno(ScalingCorrelationFunctionStronger)
$$
In the case of our 3D model, we would replace $t^n$ in
Eq.\ \(ScalingCorrelationFunctionStronger) by
$\log(t)$.  However, we know such a form will not hold very well for our data
since we are just beginning to see the expected asymptotic behavior of $L(t)$
at the
latest times reached.  Thus we shall concentrate on determining whether we
can see the weaker form of scaling implied by
Eq.\ \(ScalingCorrelationFunctionAgain).

We are also be interested in the existence of anisotropy in the correlation
function.  To investigate this, we look at the correlation function
along a few different directions.  In two dimensions, we consider it along
the lattice
axes and along the lattice diagonals.  In three dimensions, we look
along the lattice directions, the face diagonals ($[110]$ direction),
and the body diagonals ($[111]$ direction).  As an example, here is the
explicit definition we use for the correlation
function along the face diagonals in three dimensions:
$$
\vbox{
   \eqalignno{
        C_{[110]}({\sqrt 2} r) = {1\over{6N}}
	{\sum_i \Bigl[ }
	  &s({\bf r}_i) s({\bf r}_i + ({\bf {\hat x}} + {\bf {\hat y}}) r)
	+ s({\bf r}_i) s({\bf r}_i + ({\bf {\hat x}} - {\bf {\hat y}}) r)
	+ \cr &s({\bf r}_i) s({\bf r}_i + ({\bf {\hat y}} + {\bf {\hat z}}) r)
	+ s({\bf r}_i) s({\bf r}_i + ({\bf {\hat y}} - {\bf {\hat z}}) r)
	+ \cr &s({\bf r}_i) s({\bf r}_i + ({\bf {\hat z}} + {\bf {\hat x}}) r)
	+ s({\bf r}_i) s({\bf r}_i + ({\bf {\hat z}} - {\bf {\hat x}}) r)
	\Bigr]
        \ ,  &(DefineCorrelationFunction) \cr }
}
$$
where $s({\bf r}_i)$ is the value of the Ising spin at the location
${\bf r}_i$, and $r$ is a nonnegative integer. (The trivial dependence on
time has been suppressed for brevity.)

It is important to notice that our measure
of $L(t)$, Eq.\ \(MeasureOfCharacteristicLengthScale), is inversely
proportional to $C_{[100]}(0) - C_{[100]}(1)$ (using the notation for the 3D
case).  Thus, when we measure
$L(t)$ in this way, we are studying the initial
slope of the correlation function in the direction of the lattice axes.
This has important implications for the scaling
plots to be presented below:  When we plot $C_{[100]}(r,t)$ versus $r/L(t)$,
we are forcing the initial slopes of the correlation function at different
times to be
equal.  Therefore we will necessarily see good scaling near the origin.
This limits the extent to which we can determine whether or not scaling is
satisfied, particularly when we look at $C({\bf r},t)$ along the lattice axes.

For the 2D nearest--neighbor ($J_2 = 0$) Ising model,
Humayun and Bray\refto{HumanyunAndBray} have recently tested scaling
with extensive simulations on systems of size $1000^2$.  They find
that scaling is well--obeyed and furthermore that there is no evidence
for anisotropy in the scaling function (\ie scaling plots for $C({\bf r}, t)$
along the lattice axes and along the lattice diagonals fall on top of one
another).  As a check on our numerics, we have repeated this calculation on
$500^2$ systems and find the same results.\refto{Thesis}

Let's now consider scaling in the 2D model for $J_2 \ne 0$.
\Fig{scaling_2d.ps} shows results for two values of $T/J_2$.  There are
three points to be made.  First, there is little or no difference
seen for the two values of $T/J_2$.  Second, unlike the case of
$J_2 = 0$, here we can clearly see anisotropy in the correlation function.
This should not surprise us given the ``blocky'' appearance of the
configurations.   However, we believe that this anisotropy should
disappear once $L(t) \gg e^{4J_2/T}$, in which case the domain boundaries
should have a significant number of kinks in them.
For the two temperatures which we have shown in \Fig{scaling_2d.ps}, we are
always in the regime
$L(t) < e^{4J_2/T}$.  We have investigated the anisotropy at two higher
temperatures, $T = 1.8 J_2$ and $T = 3.6 J_2$.\refto{Thesis}
Particularly for the latter,
we can get out to the regime $L(t) \gg e^{4J_2/T}$, and we find the anisotropy
is indeed less pronounced,
particularly at later times.  However, the shift in the curves
with time is quite small and there is still
some detectable anisotropy even at the latest times [where $L(t) \approx 30$]
for both temperatures.

The third point concerns scaling:  When we look only along
the lattice axes, the correlation function appears to scale fairly well over
all times.  However, we have explained above why this is a rather poor test of
scaling.  When we look instead along the lattice diagonals, scaling is fairly
good only at late times, once $t^{1/2}$ growth of $L(t)$ has resumed (although,
as noted above, we would predict that the curves should
continue to shift very slowly to the right over time to reduce the
anisotropy).  At earlier times the correlation function doesn't satisfy
scaling, with the curves lying clearly to the left of those at later times.

Finally, we look at scaling in the 3D model
for $J_2 \ne 0$. \Fig{scaling_3d.ps} shows a
scaling plot for the correlation function for $T = 2 J_2$ and $T = 3 J_2$
along the lattice axes; and for only $T = 3 J_2$
along the face and body diagonals. As in two dimensions, we can clearly see
anisotropy in the correlation function.
We {\it conjecture} that, unlike what we expect in two dimensions, here
this anisotropy will remain out to all times.
Along the lattice axes, the collapse of the data appears to be quite good
(besides a little deviation upward in the tails at the latest few
times, probably as a result of finite--size effects),
and there is no perceivable dependence on the value of $T/J_2$. However,
in the diagonal directions, there is a tendency for the data at early times
(say, $t \le 650$ MC steps/spin) to lie below and to the left of
the later time data, gradually
approaching it as time increases.  For the later times, scaling is quite good.
This suggests that the scaling regime for $T = 3 J_2$ is not reached until
$t \approx 10^3$ MC steps/spin.  This is about the same time when the growth
in $L(t)$ begins to look roughly logarithmic.

The conclusion from our studies of the correlation functions is that at
late times, scaling [in the weaker sense of
Eq.\ \(ScalingCorrelationFunctionAgain)] appears to be satisfied
in two and three dimensions even when $J_2 \ne 0$.
However, our results are certainly not
sensitive enough to preclude the possibility of small deviations from scaling.
At earlier times, there are some deviations from scaling.
These are likely due to the fact that the NNN AFM bonds,
in slowing the growth of $L(t)$, also slow the approach to the scaling regime.
The one clear result of the NNN bonds is to make the
correlation function anisotropic.  This is not surprising given
the blockiness of the configurations when $J_2 \ne 0$.  Our conjecture
is that this anisotropy will remain in the
3D model out to all times, while in the
2D model the anisotropy will slowly decrease once $L(t) \gg e^{4J_2/T}$.

\references

\refis{Thesis} J.~D. Shore, Ph.D. thesis, Cornell University (1992).


\refis{WhyNoDisorder} Structural glasses have no imposed disorder:
the disorder freezes in at the glass transition.

\refis{WhatFrustration} The term ``frustration'' has many meanings.
In Section~V we discuss in what sense we consider our models
to be frustrated. Suffice it to say here that we consider them to be frustrated
in a dynamical sense. (\Ie one finds that in order for the system to lower its
energy in the long run, it must go through states of higher energy
in the short run.)

\refis{Dixon} P.~K. Dixon and S.~R. Nagel, \prl 61, 341, 1988;
P.~K. Dixon, \prb 42, 8179, 1990.

\refis{ZeroTemperatureGlauber} We generalize this to zero temperature
by taking the limit $T \to 0$ in Eqn. \(Glauber), which gives
$$
  P = \cases{0, &if $\Delta E > 0$\ ;\cr
  1/2, & if $\Delta E = 0$\ ;\cr
  1, & if $\Delta E < 0$\ .\cr}
$$

\refis{Pedagogy} The 2D model also serves an important
pedagogical purpose,
because, unlike in the examples discussed by Lai \etal [Ref.\ \cite{Lai}],
here there is only a single
barrier height ($4J_2$) in the coarsening process.  As a result, this model
is a very clean example of a class 2 system.

\refis{Glauber} R.~J. Glauber, {\sl J. Math. Phys.} {\bf 4}, 294 (1963).

\refis{Metropolis} N. Metropolis, A.~W. Rosenbluth, M.~N. Rosenbluth,
A.~H. Teller, and E. Teller, {\sl J. Chem. Phys.} {\bf 21}, 1087 (1954).

\refis{Statics} For equilibrium properties of the 2D model,
see J. Oitmaa, {\sl J. Phys. A}{\bf 14}, 1159 (1981);
D.~P. Landau and K.~Binder, \prb 31, 5946, 1985; and references therein.

\refis{Lifshitz} I.~M. Lifshitz, {\sl Zh. Eksp. Teor. Fiz.} {\bf 42},
1354 (1962) [{\sl Sov. Phys.--JETP} {\bf 15}, 939 (1962)].

\refis{AllenCahn} S.~M. Allen and J.~W. Cahn, {\sl Acta. Metall.} {\bf 27},
1085 (1979).

\refis{FrozenInTime} S. Brawer, {\it Relaxation in Viscous Liquids and
Glasses} (The American Ceramic Society, Columbus, OH, 1985).

\refis{ToOurGlassPaper} For a more detailed discussion of the slow
  dynamics in glasses and our theoretical views on the matter, see
  Ref.\ \cite{OurGlassPaper}.  That paper (and the abbreviated discussion here)
  presents an elaboration of scaling
  ideas about the glass transition suggested to us by Daniel Fisher
  [D.~S. Fisher, private communication].

\refis{OurGlassPaper} J.~P. Sethna, J.~D. Shore, and M. Huang,
{\sl Phys. Rev. B} {\bf 44}, 4943 (1991).

\refis{HowWellFit} It is not always possible to fit to the Vogel--Fulcher law
over the entire range of viscosity with one set of values for the parameters.
The controversy over the extent to which the viscosity
and other data really do obey the Vogel--Fulcher law, along with alternate
fits to the data and theories of the glass transition,
are discussed further in Ref.\ \cite{OurGlassPaper}.  See also
Ref.\ \cite{Dixon}.

\refis{PseudoEquilibrium}
We use the concept of equilibrium in a rough sense here,
since clearly a supercooled liquid is never in true equilibrium.
What interests us, however, is not the ordered solid state that would have
been formed (which we will assume is thermodynamically inaccessible), but
rather the metastable liquid state.   We say that the liquid falls out
of equilibrium when it can no longer ``equilibrate'' in the sense
of following the free energy minimum of this metastable state.

\refis{LifshitzSlyozov}  I.~M. Lifshitz and V.~V. Slyozov, {\sl J. Phys.
Chem. Solids} {\bf 19}, 35 (1961); E.~M. Lifshitz and L.~P. Pitaevskii,
{\it Physical Kinetics}
[{\it Landau and Lifshitz: Course of Theoretical Physics}, Vol. 10]
(Pergamon Press, Oxford, 1981).

\refis{Huse}  D.~A. Huse, \prb 34, 7845, 1986.

\refis{FisherAndHuse} D.~S. Fisher and D.~A. Huse, \prl 56, 1601, 1986.

\refis{Imbrie} J.~Z. Imbrie, \prl 53, 1747, 1984; {\sl Comm.\ Math.\ Phys.}
{\bf 98}, 145 (1985).

\refis{GrinsteinFernandez} G.~Grinstein and J.~F. Fernandez,
\prb 29, 6389, 1984.

\refis{Grinstein} A very readable account of the controversy, written at
a time just before it was finally settled, is given by G.~Grinstein,
{\sl J. Appl. Phys.} {\bf 55}, 2371 (1984).

\refis{Shumway} For related work on a model
which does show glassy behavior
upon slow cooling (and in fact, a correlation length which grows
only as the logarithm of the logarithm of the cooling rate), see
S.~L. Shumway and J.~P. Sethna, \prl 67, 995, 1991;
S.~L. Shumway, Ph.D. thesis, Cornell University (1991).  That model, however,
has its phase transition at zero temperature, whereas to model a glass
we are looking for such slow growth of order associated with a nonzero
transition temperature, $T_0$.

\refis{MoreOnCoarsening}   For
a more detailed  discussion, the reader is referred to
Ref.\ \cite{Thesis}.  A detailed early review of the field is provided by
J.~D. Gunton, M. San Miguel, and P.~S. Sahni, in {\it Phase Transitions
and Critical Phenomena}, edited by C. Domb and J.~L. Lebowitz (Academic,
New York, 1983), Vol. 8, p. 269.  A collection of theoretical and
experimental articles on coarsening
can be found in {\it Dynamics of Ordering Processes in Condensed Matter},
edited by S. Komura and H. Furukawa (Plenum, New York, 1988).

\refis{Amar} J.~G. Amar, F.~E. Sullivan, and R.~D. Mountain, \prb 37, 196,
1988.

\refis{MCRGIsingConserved} C. Roland and M. Grant, \prb 39, 11971, 1989.

\refis{VinalsandGrant} J. Vi\~nals and M. Grant, \prb 36, 7036, 1987.

\refis{Lai}  Z.~W. Lai, G.~F. Mazenko, and O.~T. Valls, \prb 37, 9481, 1988.

\refis{LogIsingConserved} G.~F. Mazenko, O.~T. Valls, and F.~C. Zhang, \prb 31,
4453, 1985; \prb 32, 5807, 1985; G.~F. Mazenko and O.~T. Valls,
\prb 33, 1823, 1986.

\refis{LogPotts}  S.~A. Safran, \prl 46, 1581, 1981.

\refis{EarlierPotts} M.~P. Anderson, D.~J. Srolovitz, G.~S. Grest, and
P.~S. Sahni, {\sl Acta Metall.} {\bf 32}, 783 (1984).

\refis{PottsLattice} In the case of the Potts model, whether or not such
freezing occurs depends on the lattice and on the range of the
interactions [Refs.\ \cite{NoLogPotts} and\ \cite{VinalsandGrant}].

\refis{EarlyPotts} S.~A. Safran, P.~S. Sahni, and G.~S. Grest, \prb 28, 2693,
1983; P.~S. Sahni, D.~J. Srolovitz, G.~S. Grest,
M.~P. Anderson, and S.~A. Safran, \prb 28, 2705, 1983.

\refis{NoLogPotts}  G.~S. Grest, M.~P. Anderson, and D.~J. Srolovitz,
\prb 38, 4752, 1988.

\refis{Skepticism} For example, in Ref. \cite{NoLogPotts}, Grest \etal
consider further--neighbor interactions in the 3D Potts model with the
justification that they want to avoid zero--temperature freezing due to local
lattice effects in order to better study the universal features of the domain
growth.
Our claim is that it is possible for such lattice effects to
alter the univeral features, that is, to change the asymptotic growth
law for the model.  Do we expect the nearest--neighbor 3D Potts model to have
logarithmic coarsening?  There are special configurations which
have an energy cost per unit length involved in expanding any of the domains
along that boundary.  On the other hand, it is not at all clear
that such configurations will form with sufficient probability to pin the
coarsening system.  Therefore, we are not sure what to expect, but believe it
is {\it possible} that the model does not coarsen as $t^{1/2}$.

\refis{Universality} See, \eg J. Vi\~nals and D. Jasnow,
\prb 37, 9582, 1988; C. Jeppesen, H. Flyvbjerg, and O.G. Mouritsen,
\prb 40, 9070, 1989.
Recently, however, Mouritsen and co-workers have themselves made a claim of
logarithmic coarsening in a model without randomness,
namely a 2D Ising model with diffusing (\ie annealed)
vacancies [P.~J. Shah and O.~G. Mouritsen, {\sl Phys. Rev. B} {\bf 41},
7003 (1990)].
Their claim is based upon numerical simulations of such a coarsening system.
Gregory Hassold and David Srolovitz have disputed this
claim [G.~N. Hassold and D.~J. Srolovitz, private communication],
since they have seen similar
slowing of the growth for diffusing impurities in their grain growth
(Potts model) simulations, but with a return to $t^{1/2}$ behavior at later
times.

\refis{RottmanAndWortis} C. Rottman and M. Wortis, \prb 29, 328, 1984;
{\sl Phys. Rep.} {\bf 103}, 59 (1984).

\refis{UniversalConfusion} These classes should
not be confused with the universality classes which are used to distinguish the
different scaling behaviors at long times. For example, classes 1 and 2 will
both correspond to the same universality class since their long time
scaling behavior is the same.

\refis{NoBoltzmann} Throughout this paper (except when
considering experimental data), we measure temperatures in
units of energy, that is, we set $k_B = 1$.

\refis{DifferentNotation} In Refs.\ \cite{RottmanAndWortis},
\ \cite{TilingModel}, and \ \cite{Wortis}, the authors use $T_0$ and $T_3$ to
denote the corner and edge rounding
transition temperatures, respectively.  Note that the tricritical temperature
$T_t$ discussed in Ref.\ \cite{RottmanAndWortis} is no longer believed to
exist (as explained in Ref.\ \cite{Wortis}).

\refis{Wortis} M. Wortis in {\it Chemistry
and Physics of Solid Surfaces VII}, edited by R. Vanselow and H.~F. Howe,
(Spring-Verlag, Berlin, 1988), p. 367.

\refis{FirstCommunication} In our first communication on this
work [Ref.\ \cite{Shore}], we did not fully
appreciate the connection of our work with the work on equilibrium
crystal shapes and, as a result, referred to both
these transition temperatures as ``edge roughening'' temperatures.
We called $T_{ER}$ the roughening temperature for an infinite edge, and
identified $T_{CR}$ as the roughening temperature for a finite edge (\ie an
edge which ends in a corner).

\refis{Justification} It is perhaps worth emphasizing that
our justification for the claim does not come from imagining
that the cubes we shrink are themselves equilibrium crystal
shapes.  In fact, they should not be thought of as equilibrium objects at all.

\refis{GeometricFactor} In detail, the barrier free energy is related to
the step free energy as follows:
Unlike the step free energy, the barrier free energy is calculated not per unit
length of step, but per unit length of step {\it as projected} onto one of the
lattice axes (\ie per unit length of cube edge).
We perform this calculation by summing over all
configurations without any restrictions on the average orientation of the step.
In the thermodynamic limit, we therefore calculate
the barrier free energy per unit projected length for the most probable
orientation of such a step/barrier.
The most probable orientation with respect to the crystal axes, $\theta$,
is easily computed to be
$$
\tan[\theta(T)] = {x^2 \over {(1-x) (1-x+x^2)} } \ \ ,
$$
where $x \equiv \exp(-4 J_2 / T) $.
Our barrier free energy is nothing but the step free energy for a step of
orientation $\theta$, modulo a geometric factor of $\cos(\theta)$ which
accounts for the projection.


\refis{WhyBarrierAboveToo} If we look at the Monte Carlo
data in \Fig{shrinking_cubes_0_full.ps}, we see that there is a
large temperature region above $T_{CR}$ where the free energy barrier
$F_B$ looks to be roughly $L$--independent
but still nonzero.  This may at first seem surprising.  However,
it is important to note that the fact that $f_B$ goes to
zero at $T_{CR}$ does not mean the free energy
barrier, $F_B$, for finite cubes goes to zero there.  It means
only that $F_B$ becomes independent of system size. (Or, possibly, that
it grows with $L$ but more slowly than in direct proportion! In
\Fig{shrinking_cubes_0_full.ps}, there does
actually appear to be some small $L$--dependence in the barrier even above
$T_{CR}$.) In fact, we know that the rate to flip a corner spin is
$$
        {e^{-12J_2/T}\over{1 + e^{-12J_2/T}}}
        \ ,
$$
and thus that there is an energy barrier to perform this elementary process
regardless of whether we are below or above $T_{CR}$.  (A similar
argument for the 2D model explains why we see a barrier of
$4 J_2$ even though the roughening temperature for this model is $T_R = 0$.)
Why then does the time to flip the edge of a cubic domain appear to
become completely temperature--independent (or perhaps even rise a bit
with $T$) at $T \approx J_1$?  It is conceivable that this could
occur because at the edge rounding transition temperature
($T_{ER} \approx J_1$), it becomes favorable to depin a step from the
cube edge without having to work in from the corners.
However, we have not been able to make such an association rigorous and
it seems more likely that this apparent temperature independence is
simply a quirk
of our criterion for determining when the spins along an entire edge
have flipped (see Subsection~A), since this criterion becomes
suspect at temperatures high enough that the equilibrium state itself would
have a significant number of thermal excitations.

\refis{MoreDetailsOnThis}  More detail on this point can be found in
Ref.\ \cite{Thesis}.

\refis{CruderEstimate} In Ref.\ \cite{Shore}, we
reported $T_{CR} \approx 9$---$10 J_2$.  This cruder
estimate was obtained from calculating $f_B$ to only second
order in $e^{-4J_2/T}$.

\refis{Burton}
W.~K. Burton, N. Cabrera, F.~C. Frank, {\sl Phil. Trans. Roy. Soc.
(London)} {\bf 243A}, 299 (1951).

\refis{Macroscopic} The term
``macroscopic'' is used to emphasize the fact
that the ECS is defined as the shape of a crystal in the thermodynamic
limit (that is, as the ratio of the crystal size to the lattice constant
goes to infinity).

\refis{vanBeijeren} H. van~Beijeren and I. Nolden, in {\it Structure and
Dynamics of Surfaces, II}
[{\it Topics in Current Physics}, Vol. 43],
edited by W. Schommers and P. vonBlanckenhagen
(Springer-Verlag, Berlin, 1987), p. 259.

\refis{Weeks} J.D. Weeks, G.H. Gilmer, and H.J. Leamy, \prl 31, 549, 1973.

\refis{TimeToShrink}
In the simulation,
the first edge is considered to have flipped
when the $J_1$ energy first falls below its initial value.  We choose to show
this time rather than the time to shrink the cubic domain entirely
because it is
easier to compare with analytic calculations.  The time to shrink the
entire domain behaves similarly (See Ref.\ \cite{Shore}, Fig.~1).
At any rate, the time to flip an edge is certainly a lower
bound on the time to shrink the entire cube, and is in fact quite a good
bound in the low temperature limit where the time to shrink the cube is
dominated by the time to flip the first edge.

\refis{MightExpect} Whether
we allow the edge to be eaten away from one or both corners does not matter
in the thermodynamic limit.---Eq.\ \(FreeEnergyBarrier4) also simplifies to
Eq.\ \(FreeEnergyPerLength2).

\refis{HighOrder} This should be reasonable because the error made here will
be of high order in $e^{-4J_2/T}$. Of course, it is exact in
the thermodynamic limit.

\refis{GrubbyDetails} The factor of 2 accounts for the fact that for each
edge, spins
can be flipped on either one of the two adjoining faces.  The -1 corrects
for the double counting of the lowest barrier configuration
[\Fig{cubes_galore}(a)].

\refis{DifferentRougheningTemperatures} More generally, facets with
different symmetry orientations can have
different roughening temperatures. For the simple cubic Ising model
with $J_2 = 0$ the only stable facets are the ones perpendicular to the
crystal axes (\eg [100]) and there is only one $T_R$.

\refis{InPractice} In practice, we compute step
free energies without explicitly specifying the step orientation.  This means
that we sum over all possible orientations and thus get the step free energy
for the step with the most probable orientation [Ref.\ \cite{MarkThesis}].

\refis{Shore} J.~D. Shore and J.~P. Sethna, \prb 43, 3782, 1991.

\refis{JapanConferencePaper} J.~D. Shore, J.~P. Sethna, M. Holzer, and
V. Elser, to appear in ``Slow Dynamics in Condensed Matter,'' Proceedings of
the 1st Annual Tohwa University International Symposium, Fukuoka, Japan
(American Institute of Physics, 1992).

\refis{MarkThesis} M. Holzer, Ph.D. thesis, Simon Fraser University (1990).


\refis{Tubes} See, \eg Ref.\ \cite{VinalsandGrant}.
They discuss a two--dimensional system, where one can have strips infinite in
one dimension.  In a three--dimensional system, one can have both
tubes (infinite in one dimension) and slabs (infinite in two dimensions).

\refis{Averaging} Two technical details about averaging $L(t)$ over
several runs:
First, we define the standard error in $L$ by
${\sqrt{(\langle L^2 \rangle - \langle L\rangle^2)/N_R}}$,
where $\langle ...\rangle$ denotes averaging and $N_R$ is the number of runs
averaged over.  Second, for these coarsening results, we have actually
computed the average and standard error in the quantity
$1/L$ rather than in $L$ itself.  The motivation for doing this is to prevent
one run in which $L$ gets very large from dominating the average (after all,
as it is defined  in Eq.\ \(MeasureOfCharacteristicLengthScale),
we can have $L \to \infty$).  For the results
presented in this paper, the difference between the two ways of averaging is
very small except once the finite--size effects are obviously dominating.
Even then, the difference is less than the size of the error bars.

\refis{ThirdParameter} Such a form does introduce a third parameter,
$L_0$, obtained as a constant of integration.  However, we perform the fits
keeping the parameter $L_0$ fixed, and the fits are good as long as it is
fixed at
some reasonable value (\eg $L_0 = 1$ or $3$ work equally well).  Furthermore,
we find that fits to the form derived from integrating the equation analogous
to Eq.\ \(TrueEquation) but without the $1/L$ factor (and again
with $L_0$ fixed to 1 or 3), give fits
almost indistinguishable from those to the form $L(t) = a \, \log(t/t_0)$.
This suggests that it is not the introduction of a third parameter, but rather
the more accurate portrayal of the physics, which leads to the superior fits
obtained from Eq.\ \(TrueEquation).

\refis{Percolation} D. Stauffer, {\sl Introduction to Percolation Theory}
(Taylor and Francis, London, 1985).

\refis{HumanyunAndBray} K. Humayun and A.~J. Bray, {\sl J. Phys. A} {\bf 24},
1915 (1991).

\refis{BlockyNotEnough} It is worth emphasizing that the observation
that the configurations look blocky does not in itself imply that the growth is
logarithmic.  (After all, the configurations in the 2D model look similarly
blocky out to the times studied.)  It is this blockiness coupled with the
argument that such blocky
configurations will have barriers which grow with the length scale
that supports our logarithmic growth hypothesis.
(Whereas, in two dimensions, blocky configurations do not have barriers
which grow with length scale.  Moreover, we expect the blockiness in two
dimensions to decrease once $L(t)$ gets very large [see Appendix~D].)

\refis{AmarAndFamily} J.~G. Amar and F. Family, {\sl Bull. Am. Phys. Soc.}
{\bf 34}, 491 (1989); private communication.

\refis{PottsExample} For example, for the Potts model on a triangular lattice,
one can identify metastable configurations, that is, configurations in which
there are energy barriers to any possible state change.
However, in the coarsening process, these metastable
configurations are not nucleated in sufficient number to pin the entire
system and the model coarsens (\ie it does not freeze) at $T = 0$
[Refs.~\cite{VinalsandGrant} and \cite{EarlyPotts}].

\refis{OtherMeasures} D. Chowdhury, M. Grant, and J.D. Gunton,
\prb 35, 6792, 1987.

\refis{GroundStateOK} Ref.\ \cite{OtherMeasures} notes the importance of using
the equilibrium energy $E_{eq}$, rather than the ground state energy $E_0$, in
Eq.\ \(MeasureOfCharacteristicLengthScale), in order to correct for thermal
fluctuations.  However, here we will be working in the limit of $T \ll T_C$,
so to an excellent approximation $E_{eq} = E_0$.

\refis{Random}  G.S. Grest and D.J. Srolovitz, \prb 32, 3014, 1985.

\refis{CorrectGroundState}  Alternatively, one might be tempted to argue that
we should
change our definition of $L(t)$ [Eq.\ \(MeasureOfCharacteristicLengthScale)]
by defining ${E^{\rm NN}_0 \equiv (-3 N + 6 {\cal L}^2) J_1}$, which is the
lowest value of the nearest--neighbor energy that the system can have with
antiperiodic boundary conditions.  However, this is {\it not} correct since we
want
to define $L(t)$ as being inversely proportional to the total domain perimeter,
regardless of our boundary conditions.  For any remaining skeptics, we note
that if we do use ${E^{\rm NN}_0 \equiv (-3 N + 6 {\cal L}^2) J_1}$ in
Eq.\ \(MeasureOfCharacteristicLengthScale), we find that the simulation results
for $L(t)$ have a strong dependence on the system size even at early times,
whereas, using ${E^{\rm NN}_0 \equiv -3 N J_1}$ gives results independent of
system size (until times late enough that finite--size effects are obviously
important).


\refis{OtherPowers} What determines $r$ is the value of the exponent $n$
which would {\it naively} describe the coarsening system
(\ie, $n = 1/3$ for conserved order parameter, or $1/2$ for nonconserved
order parameter) and the value of the
exponent $m$ describing the behavior of $f_B$ as the ordering transition is
approached from below: $f_B(T) \sim \> (T_{CR} - T)^m$.
If one generalizes the argument we gave to the case of general $m$ and $n$,
one finds $r = m n / (n + m)$.  However, it is unclear to us whether or not
this would hold in the case of a second order transition, where there are
fluctuations on all length scales in the equilibrium system as one approaches
the transition.

\refis{FisherRFIM} D.~S. Fisher, {\sl Phys.\ Rev.\ Lett.} {\bf 56}, 416 (1986).

\refis{AsymptoticallyCorrect}  Numerical integration of \(TrueEquationTiling)
[for the cooling schedule defined by \(TimeAnneal)] confirms that this
differential equation indeed yields $L(T=0,\Gamma) \sim \Gamma^{-1/4}$
as $\Gamma \to 0$,
thus providing a conclusive check on the approximations we made in
deriving \(LAnneal4) and \(ExponentAnneal).

\refis{ExampleOfAttempt} For example,
in previous work on coarsening [Ref.\ \cite{OtherMeasures}], it has been
suggested that an approximate way to correct for the effects of thermal
fluctuations is to use the equilibrium energy $E_{eq}$, rather than the
ground state energy $E_{0}$, in definition of $L$
[Eq.\ \(TilingMeasureOfCharacteristicLengthScale)].  However, below $T_{CR}$,
the tiling model is thermodynamically frozen with
$E_{eq} = E_0$.  Out of equilibrium, the thermal fluctuations
are dependent on the characteristic length scale $L(t)$ itself. Thus, there
seems to be no such obvious way to correct our measure of the characteristic
length scale.

\refis{RandomInitialTiling} By infinite temperature, we mean that
$T/J_2 \to \infty$ keeping $T/J_1 = 0$.  That is, the initial state must
be a tiling.  Such an initial tiling cannot be produced here
simply by choosing a random spin configuration, since such a
configuration will not, in general, correspond to a tiling at all.
We thus prepare the $T = \infty$ initial state as follows:
First, we tile the plane with the elementary hexagons (\ie the shaded objects
in \Fig{tiling_dynamics}), with each hexagon chosen randomly to
be in one of the two possible orientations.  Such a tiling is not
a $T = \infty$ state, however, since its entropy is clearly not
maximized.  It is necessary to randomize this tiling by choosing spins
randomly and flipping them (without regard to the energy cost) so long as
three of their six nearest neighbors are aligned,
during which time the average energy of the system decreases.
We perform such randomization for at least $100$ MC steps/spin, by which
time the average energy has levelled off.  The resulting tiling is our
initial, $T = \infty$ state.

\refis{CorrectGroundState2} Note that we do {\it not} define
$E_0 \equiv (- N + 8 \sqrt{N}) J_2$,
which is the lowest energy state that can be reached by this system (given
the boundary conditions and the restriction that the number of each type
of tile remains equal).  See Ref.\ \cite{CorrectGroundState}.

\refis{Veit} Our study of this model was initiated at the suggestion of
Veit Elser [V.~Elser, private communication].

\refis{TilingModel} A.-C. Shi and M. Wortis, \prb 37, 7793, 1988.

\refis{Jayaprakash} C. Jayaprakash and W.~F. Saam, \prb 30, 3916, 1984.

\refis{Blote} H.~W.~J. Bl\"ote and H.~J. Hilhorst, {\sl J. Phys. A} {\bf 15},
L631 (1982); B. Nienhuis, H.~J. Hilhorst, and H.~W.~J. Bl\"ote,
{\sl J. Phys. A} {\bf 17}, 3559 (1984).  These authors considered a model
with the same tiling configurations but with a different Hamiltonian (\ie
different energies assigned to the configurations).

\refis{EqualTiles} The average orientation of
the interface is controlled by the relative number of each type of tile.
We will only be interested in the case of equal numbers of each tile;
that is, in a [111] interface.

\refis{ItFollows} The boundary between unlike tiles corresponds to a bend in
the interface, which costs an energy of $2 J_2$ per unit length
(See Appendix~B).

\refis{Why} The point is that,
from the three--dimensional viewpoint, digging a hole in, or building upon, a
flat ([100], [010], or [001]) section of interface will
necessarily produce some hidden interface, as viewed from the [111]
direction.

\refis{WideLiterature} Reconstructions of this type have an extensive
literature [Refs.\ \cite{Herring} and\ \cite{ThermalFaceting}]
and are referred to by several different names, including
``hill-valley reconstruction,'' ``Herring reconstruction,''
``thermal etching,'' and ``thermal faceting.''

\refis{CubicalProjections} For the tiling model, the activation barriers
can be studied by looking at the time to shrink a ``cubical projection''.
Such a Monte Carlo study has been conducted [Ref.\ \cite{Thesis}]
and yields results qualitatively
similar to those shown in \Fig{shrinking_cubes_0_full.ps}.

\refis{InEffect} In effect, we are asking the question, ``If
$L(t)$ resumed a growth law of $t^{1/3}$ at times just after we stopped our
simulations, then what would we conclude were the heights of the largest
free energy barriers which the system had to cross during the coarsening
process?''


\refis{Heyraud} J.~C. Heyraud and J.~J. M\'etois, {\sl J. Cryst. Growth}
{\bf 84}, 503 (1987).

\refis{Knoppik} D. Knoppik and A. Losch, {\sl J. Cryst. Growth},
{\bf 34}, 332 (1976); D. Knoppik and F.-P. Penningsfeld,
{\sl J. Cryst. Growth}, {\bf 37}, 69 (1977).

\refis{GrainGrowth} Summaries of the experimental literature on
grain growth can be found in Ref.\ \cite{LatestPotts} (table 2);
Ref.\ \cite{EarlierPotts};
V. Randle, B. Ralph, and N. Hansen, in {\it Annealing Processes ---
Recovery, Recrystallization, and Grain Growth}, edited by N.~Hansen,
D. Juul Jensen, T. Leffers, and B. Ralph (Ris\o National Laboratory, Ris\o,
1986), p. 123; H. Hu and B.~B. Bath, {\sl Metal. Trans.} {\bf 1}, 3181 (1970).

\refis{LatestPotts} M.~P. Anderson, G.~S. Grant, and D.~J. Srolovitz,
{\sl Phil. Mag. B} {\bf 59}, 293 (1989).

\refis{Herring} C. Herring, \pr 82, 87, 1951; C. Herring in {\it Structure
and Properties of Solid Surfaces}, edited by R. Gomer and C.~S. Smith
(University of Chicago Press, Chicago, 1953), p.~5.

\refis{ThermalFaceting} For reviews, see A.~J.~W. Moore, in {\it Metal
Surfaces: Structure, Energetics, and Kinetics} (American Society for Metals,
Metals Park, OH, 1963), p.~155; M. Flytzani--Stephanopoulos and L.~D. Schmidt,
{\sl Prog. Surf. Sci.}, {\bf 9}, 83 (1979).

\refis{ExperimentalDifficulties}
For a summary of the experimental difficulties faced
in the study of equilibrium crystal shapes, see, \eg Ref.\ \cite{Wortis}.

\refis{LargeBody} In addition to the experiments on coarsening of
the [111] interface of NaCl, there is also a large body of related
experimental work in the materials science literature on
other systems in which an unstable interface coarsens by some sort of
hill--valley reconstruction into pieces of interface of allowed
orientations [Ref.\ \cite{WideLiterature}].
Unfortunately, the field remains somewhat murky, with
many of the experimental results in contradiction with one another.  In
particular, it is not clear how close to equilibrium the surfaces are in many
of the experiments.  This has prompted the assertion that
the mechanism by which many of the observed reconstructions occur could have
more to do with the dynamics of evaporation than with equilibration by
minimization of free
energies.  Even assuming the latter mechanism, it's hard to know
whether these systems are below $T_{CR}$ or only below $T_{ER}$, since
the equilibrium crystal shapes have not been determined for these materials.
(Most likely, the equilibrium crystal shapes are quite complicated,
with several different roughening and rounding
transitions for different facets, edges, and corners.)
Below $T_{ER}$, a [110] interface should reconstruct into
pieces of [100] and [010] interfaces, but we do not necessarily expect this
coarsening to be slow.


\refis{HolzerandWortis} M. Holzer and M. Wortis, \prb 40, 11044, 1989.


\refis{Bortz}  A.~B. Bortz, M.~H. Kalos, and J.~L. Lebowitz, {\sl J. of Comp.
Phys.} {\bf 17}, 10 (1975).  See also K. Binder in {\it Monte Carlo Methods
in Statistical Physics}, 2nd edition,
edited by K. Binder (Springer--Verlag, Berlin, 1986), Section~1.3.1.

\refis{FoundError} In fact, the comparison
initially turned up a systematic difference between the two methods.  The
difference turned out to be due to using a random number generator with a small
cycle. (It chose one of only 714025 different values.)  This had
a noticeable effect on the standard MC method because it made it
highly improbable that the same spin would be chosen for updating in relatively
quick succession.  The effect of this could be detected even when the number
of sites was only 27000, less than $4\%$  the number of values returned by the
generator! A different random number generator with a much larger number of
possible values
eliminated this discrepancy.  This illustrates the importance of comparing
different random number generators (and perhaps even different MC algorithms)!

\endreferences

\vfill\eject
\figurecaptions

{\bf Fig. 1:} {\bf (a)} A square domain of ``up'' spins in a system of
``down'' spins, in two dimensions.
The next--nearest--neighbor bonds introduce an energy barrier of $4 J_2$ to
flipping a corner spin (dark gray). As a result, it takes a time of
order $e^{4 J_2/T}$ to shrink the entire domain.
{\bf (b)} A cubic domain in three dimensions.
(Here, for clarity, most of the individual spins have not been shown.)
There is an energy barrier of $12 J_2$ to flip a corner spin (dark gray) and
a barrier of $4 J_2$ to flip each spin along the edge (light gray).  Thus,
unlike in two dimensions,
the total barrier to flip all the spins along an edge is
proportional to the linear size, $L$, of the domain; and the time to shrink
the domain is now exponential in $L$.

{\bf Fig. 2:} {\bf Shrinking of a Square Domain in Two Dimensions.}
This shows the times to shrink a square domain versus $J_2/T$, as obtained from
Monte Carlo simulations
with $J_1/J_2 = 6$.  The three sets of Monte Carlo data are for
domains of size $L =$ 4, 6, and $8$ (bottom to top).
Each point is an average over 900 runs with standard error smaller than the
symbol size.  Lines are one parameter fits to
Eq.\ \(Shrink2d) using the ten lowest temperature data points.
(If we instead make the free energy barrier $F_B$ a free
parameter too,
we find $F_B/J_2 = 4.00 \pm 0.01$, $3.99 \pm 0.01$, and $3.99 \pm 0.01$ for
$L = 4, 6,$ and $8$, respectively.)
This confirms that shrinking a square domain
involves activation over barriers of height $4 J_2$, independent of
domain size.

{\bf Fig. 3:} {\bf Shrinking of a Cubic Domain in Three Dimensions.}
Shown is the time to flip the first edge of a cubic domain of size $L$
versus $J_2/T$ from Monte Carlo simulations for $J_1/J_2 = 100$.
($J_1/J_2 = 6$ yields results essentially indistinguishable from this
for $T < 4 J_2$.  At higher temperatures the results for these two
values of $J_1/J_2$ differ markedly.)
Each point is an average over 900 or 1600 runs with standard error smaller
than the symbol size.
As expected, the slope of the data increases with cube size.
For each value of $L$, except $L = 24$, we show one parameter fits to
Eq.\ \(Shrink3d) using the 7 to 10 lowest temperature points.  For small $L$,
the fits are quite good; but for $L = 8$ and $12$, the fits are clearly
inadequate.

{\bf Fig. 4:}
Various configurations of a cube with all but one spin along the top edge
flipped.  {\bf (a)} shows the configuration with the lowest possible
energy barrier of $4 J_2 (L+1)$. {\bf (b)} shows an additional row flipped.
This has an energy of $8 J_2$ higher than (a).  {\bf (c)} shows two additional
rows flipped.
{\bf (d)} shows an edge being ``eaten'' away from both corners.
Finally, (e) and (f) show two configurations
not counted in our computation of the free energy barrier.  {\bf (e)} is
neglected because
it is not a barrier configuration --- Flipping the last spin in the second
row lowered the energy by $4 J_2$.  {\bf (f)} is not included because spins are
flipped in more than one layer.  It is thus outside our approximation of
considering only those barrier configurations in which just one layer is being
peeled away from each corner.

{\bf Fig. 5:} Energies of configurations involved in approximating the
free energy barrier $F_B$.  If we let $m_i$
label the difference in the step height between successive columns, as shown,
then the energy associated with the $i$th column is given by
$E(m_i) = 4 J_2 (m_i+1)$
for $m_i > 0$ and by $E(0) = 0$.  (Of course, there will be an overall
additional energy of $4J_2 (L+1)$ from flipping the spins along the top row.)

{\bf Fig. 6:} The same simulation results as
\Fig{shrinking_cubes_0_full.ps}, but with improved theoretical forms.
The solid curves are
the theoretical form of Eq.\ \(TimeFromFreeEnergyBarrier) with
$\Gamma_0$ given by Eq.\ \(Gamma0Average) and the free energy barrier
$F_B$ given by Eq.\ \(FreeEnergyBarrier7). The
dotted curves use the cruder estimate for $F_B$ given by
Eq.\ \(FreeEnergyBarrier4), which does not include configurations where the
edge is eaten from {\it both} corners.  (In both cases there are no free
parameters.) Note that in order to get essentially perfect agreement,
it is necessary to include configurations in which the the edge is ``eaten''
away from both corners, even though (as discussed in Subsection~II.C) such
configurations do not change the expression for $F_B/L$ in the thermodynamic
limit $L \to \infty$.

{\bf Fig. 7:}
The free energy barrier per unit edge length, $f_B$, versus temperature $T$, as
determined from Eq.\ \(FreeEnergyPerLength2).  The temperature $T_{CR}$ at
which $f_B = 0$ is shown by a dotted line.

{\bf Fig. 8:}  Qualitative thermal evolution of the
equilibrium crystal shape of the Ising model with weak next--nearest--neighbor
antiferromagnetic bonds (after  Ref.\ \cite{RottmanAndWortis}).
For $T < T_{CR}$, the crystal has flat faces with macroscopically sharp edges
and corners.  Above $T_{CR}$, the crystal rounds near the corners;
but up to a temperature of $T_{ER}$, at least
part of the crystal edge remains sharp.  Above $T_{ER}$, there are no sharp
edges but there are still flat [100] faces.  Finally, at the roughening
temperature $T_R$, the crystal becomes completely rounded.

{\bf Fig. 9:} {\bf Coarsening in Two Dimensions.}
This shows the growth of the characteristic length scale $L(t)$ for the
2D model following a quench from infinite temperature to
the given temperature $T$.
For the $J_2 \ne 0$ runs, we have chosen $J_1/J_2 = 6$.  We have averaged over
30 or 40 runs with error bars showing the standard error.
Numbers in parentheses give system sizes.  (On this log--log plot,
some straight lines of slope 1/2 are shown to guide the eye.)
For $J_2 = 0$, we see the expected $t^{1/2}$ growth law at all times.
For $J_2 \ne 0$, the system initially coarsens on very short length scales,
but then gets stuck. Little further coarsening occurs
on time scales shorter than $t = e^{4J_2/T}$ (marked by arrows).  However,
if we look at the coarsening on times scales greater than this, the energy
barriers can be crossed and the length scale grows as $t^{1/2}$.

{\bf Fig. 10:} {\bf Coarsening in Three Dimensions.}
This shows the growth of the characteristic length scale $L(t)$ for the
3D model following a quench from infinite temperature.
Each set of Monte Carlo data is from an average over 10---20 runs with error
bars giving the standard error.  (The scatter in the data is much less than
the error bars would suggest because the error is strongly correlated in time.)
For $T = 8 J_2$, we've chosen $J_1/J_2 = 50$, while
for the rest (excluding $J_2 = 0$, of course), $J_1/J_2 = 6$.
For $T = 2$, 3, and $4J_2$, the solid curves show two--parameter
fits at late times (over the interval for which the curve is shown)
to the form $L(t) = a \, \log(t/t_0)$.
Arrows for $T = 0.75 J_2$  indicate the times at which
$t = e^{4J_2/T}$ and $e^{8J_2/T}$.

{\bf Fig. 11:} A comparison of the spin configurations in one layer (a slice)
of an $80^3$ coarsening system for $J_2 = 0$ and $J_2 \ne 0$.
(a) and (b) show configurations for $J_2 = 0$, while (c) and (d) show
configurations for $T = 3 J_2$.  Spin up and spin down are denoted by black
and white squares, respectively.  The times have been chosen so that
$L(t)$ is approximately equal to $6.3$ in (a) and (c),
and to 9.5 in (b) and (d). As expected,
the configurations for $J_2 \ne 0$ show a much stronger preference
for flat boundaries aligned along the lattice directions.

{\bf Fig. 12:} {\bf A sample configuration for the til\-ing mod\-el.}
Viewed from the [111] direction, an interface in
a 3D model can be represented by a tiling of
the plane by rhombi of three orientations, provided that the interface has
no overhangs when thus viewed.
(We have shaded the three types of rhombi
differently in order to distinguish them more easily and enhance the
3D perspective.)  If we assign an energy of $2 J_2$ to each
unit length of boundary between the different types of tiles,
then the energetics of this model matches that of the 3D
Ising model with next--nearest--neighbor antiferromagnetic bonds.

{\bf Fig. 13:} An example of an interface configuration which cannot
be represented by the tiling model. Part of the interface is hidden
from view (\ie there is an ``overhang'') and this results in some incomplete
(triangular) tiles.

{\bf Fig. 14:} The elementary dynamical move in the tiling model.
In the 2D spin representation,
the move consists of flipping a spin which has exactly 3 of its 6
nearest--neighbors aligned.  In tiling language, it consists of a
rotation of an elementary hexagon (shaded) by $60^\circ$.  From a
3D perspective, we see that it represents an elementary
cube either added or taken away.

{\bf Fig. 15:} Snapshots of the coarsening of the tiling model
at various times $t$ (measured in MC steps/spin) following a
quench from infinite temperature to $T = 3 J_2$. This system has $120^2$ sites.
Only the boundaries between domains of equally oriented tiles are shown.

{\bf Fig. 16:} {\bf Coarsening in the Tiling Model.}
This log--log plot shows the growth of the characteristic length scale $L(t)$
for the tiling model following a quench from infinite temperature.
For $T = {2\over{3}}J_2$, we study a $60^2$ system and average over
12 runs.  For the others we study a $120^2$ system and average over
89---170 runs.
Arrows for $T = {2\over{3}}J_2$  indicate the times at which
$t = e^{4J_2/T}$, $e^{8J_2/T}$, and $e^{12J_2/T}$.  The solid line has
a slope of $1/3$.

{\bf Fig. 17:} {\bf Coarsening in Tiling Model: Long times.}
Here we show the growth of $L(t)$ for
two runs at $T = 4 J_2$ and four runs at $T = 3 J_2$ which
go out to very long times.  The solid lines are fits to the logarithmic form
$L(t) = a \, \log(t/t_0)$ using the last decade of the Monte Carlo data in
\Fig{coarsening_tiling_loglog.ps} ($t = 10^6$--$10^7$ MC steps/spin for
$T =  3 J_2$, and $t = 10^5$--$10^6$ MC steps/spin for $T = 4 J_2$).  The
dotted lines are fits to the power law form $L(t) = a\, t^n$ using the same
data.  (The value obtained for the exponent $n$ is $0.09$ for $T = 3 J_2$,
and $0.12$ for $T = 4 J_2$.)

{\bf Fig. 18:} Snapshots of the tiling model (system size = $120^2$)
as it coarsens for $T = 3J_2$ and $T = 5J_2$, both
at $t = 10^4$ MC steps/spin.  The boundaries between domains are shown.
Due to the thermal fluctuations along the boundaries, the length scale $L$
as measured by
Eq.\ \(TilingMeasureOfCharacteristicLengthScale) is slightly smaller for
$T = 5J_2$ than for $T = 3 J_2$, even though the system at $T = 5 J_2$ is
clearly considerably coarser.

{\bf Fig. 19:} The growth of the characteristic length scale $L$ for the
tiling model during cooling, for various cooling rates $\Gamma$.
Each set of Monte Carlo data is from an average over
10 runs with error bars giving the standard error.  The initial configuration
for each run is random ($T = \infty$) and then the system is quenched to
either $T = 8J_2$ or $10J_2$, at which point the slower cooling is begun.
The transition temperature $T_{CR}$ is marked by the dotted line. (The
slowness of the rise in $L$ just below $T_{CR}$ is a consequence
of thermal fluctuations causing the characteristic length scale
to be underestimated.)

{\bf Fig. 20:} Log--log plot of the length scale $L$ at $T = 0$ as a function
of the cooling rate $\Gamma$.  A line of slope $-1/4$ has been drawn
suggestively through the data at the lowest cooling rates.

{\bf Fig. 21:} Two examples of interfaces between domains.
We can compute the energies of these interfaces
by associating an energy of $E_p = 2J_1-8J_2$ with each plaquette
of interface (shaded) and an energy of $E_b = 2 J_2$ with each unit length
of bend (bold solid lines) in the interface.  For example, the interface in (b)
has 2 more plaquettes and 14 more bends than the interface in (a).  Therefore,
the energy of (b) is $4 J_1 + 12 J_2$ higher than the energy of (a).

\def\axischar#1{\hbox{\hskip2pt\lower0.4ex\hbox{%
\psfig{figure=#1.ps,height=10pt,width=10pt}%
}\hskip2pt}}
{\bf Fig. 22:} {\bf Scaling and Anisotropy in Two Dimensions.}  This shows a
scaling collapse of the correlation function, $C(r,t)$,
for the 2D Ising model.  The collapse is achieved by
plotting $C(r,t)$ versus $r/L(t)$.  Results are given for
$T = 0.42 J_2$ and $0.72 J_2$ from 30 runs each on a
$400^2$ system.  The correlation function along both
the lattice axes ($\axischar{cr}$) and the diagonals ($\axischar{sq}$)
is plotted.  [$C(r,t)$ is shown at 18 times
($10 \le t \le 2\times10^6$ MC steps/spin)
and 12 times ($10 \le t \le 2\times10^4$ MC steps/spin)
for $T = 0.42 J_2$ and $0.72 J_2$, respectively.
The late--time data along the diagonals
($t \ge 2\times10^5$ MC steps/spin for $T = 0.42 J_2$, and
$t \ge 2400$ MC steps/spin for $T = 0.72 J_2$)
are shown by larger squares than earlier time data.]
There is clearly anisotropy in the correlation function at these temperatures
out to the lastest times studied.  See text for a discussion of scaling.

{\bf Fig. 23:} {\bf Scaling and Anisotropy in Three Dimensions.}  This shows
a scaling collapse of the correlation function, $C(r,t)$,
for the 3D model along
the lattice axes (\axischar{cr}), face diagonals (\axischar{sq}), and body
diagonals (\axischar{pl}), for $T = 3 J_2$ in an
$80^3$ system, averaged over 20 runs.  The collapse is achieved by
plotting $C(r,t)$ versus $r/L(t)$.  Results at
14 times ($10 \le t \le 80000$ MC steps/spin) are included.
Also included for the case of the lattice axes are results for $T = 2 J_2$,
in an $80^3$ system averaged over 10 runs, at 18 times ($10 \le t \le
2\times10^6$ MC steps/spin).
Anisotropy in the correlation function is clearly apparent.  See text for a
discussion of scaling.

\vfill\eject

\endfigurecaptions

\nopageno
\font\bigmathb=cmmi10 scaled \magstep3
\font\scriptbigmathb=cmmi7 scaled \magstep3
\font\scriptscriptbigmathb=cmmi5 scaled \magstep3
\font\bigmatha=cmr10 scaled \magstep3
\font\scriptbigmatha=cmr7 scaled \magstep3
\font\scriptscriptbigmatha=cmr5 scaled \magstep3
\font\bigmathc=cmsy10 scaled \magstep3
\font\scriptbigmathc=cmsy7 scaled \magstep3
\font\scriptscriptbigmathc=cmsy5 scaled \magstep3
\textfont2 = \bigmathc
\scriptfont2 = \scriptbigmathc
\scriptscriptfont2 = \scriptscriptbigmathc
\textfont1 = \bigmathb
\scriptfont1 = \scriptbigmathb
\scriptscriptfont1 = \scriptscriptbigmathb
\textfont0 = \bigmatha
\scriptfont0 = \scriptbigmatha
\scriptscriptfont0 = \scriptscriptbigmatha


\FigureComposite{square_and_cube}
{Shrinking square and cubic domains}
{\topinsert}
{
\centerline{
{\psfig{figure=\figdir/square.ps,width=5.5cm}}
{\hskip 2.0cm}
{\psfig{figure=\figdir/cube.ps,width=5.5cm}}
}
\centerline{\sectionfont (a) {\hskip 5.8cm} {(b)}}
} Figure 1


\Figurepage{shrinking_squares_new.ps}{Times to shrink a square domain}
{18.0cm}{18.0cm} Figure 2

\Figurepage{shrinking_cubes_0_new.ps}{Times to flip the spins along an edge
of a cubic domain}
{18.0cm}{18.0cm} Figure 3

\FigureComposite{cubes_galore}
{Configurations for computing the energy barrier to flip an edge}
{\topinsert}
{
\centerline{
{\psfig{figure=\figdir/cube_edge.ps,width=4.3cm}}
{\hskip .5cm}
{\psfig{figure=\figdir/cube_edge2.ps,width=4.3cm}}
{\hskip .5cm}
{\psfig{figure=\figdir/cube_edge4.ps,width=4.3cm}}
}
\centerline{\sectionfont (a) {\hskip 4.0cm}  {(b)} {\hskip 4.0cm} {(c)}}
\bigskip\medskip
\centerline{
{\psfig{figure=\figdir/cube_edge5.ps,width=4.3cm}}
{\hskip .5cm}
{\psfig{figure=\figdir/cube_edge3.ps,width=4.3cm}}
{\hskip .5cm}
{\psfig{figure=\figdir/cube_edge6.ps,width=4.3cm}}
}
\centerline{\sectionfont (d) {\hskip 4.0cm}  {(e)} {\hskip 4.0cm} {(f)}}
} Figure 4

\FigureIgnoreHeight{cube_for_energy.ps}
{Calculating the energy of a configuration}
{8.0cm} {8.0cm} Figure 5

\Figurepage{shrinking_cubes_2_new.ps}{Better approximations to time to flip
a cube edge} {18.0cm}{18.0cm} Figure 6

\FigureIgnoreHeight{free_energy_per_unit_length.ps}
{Free energy barrier per unit length of edge} {14.0cm} {14.0cm} Figure 7

\vfill\eject

\FigureComposite{EquilibriumCrystalShape}
{Equilibrium crystal shapes}
{\topinsert}
{
\centerline{
{\psfig{figure=\figdir/cube1.ps,width=5.0cm}}
{\hskip .5cm}
{\psfig{figure=\figdir/cube2.ps,width=5.0cm}}
}
\centerline{$T < T_{CR}$ {\hskip 3.0cm} {$T_{CR} < T < T_{ER}$}}
\bigskip\medskip
\centerline{
{\psfig{figure=\figdir/cube3.ps,width=5.0cm}}
{\hskip .5cm}
{\psfig{figure=\figdir/cube4.ps,width=5.0cm}}
}
\centerline{$T_{ER} < T < T_R$ {\hskip 3.0cm} {$T > T_R$}}
}
Figure 8


\Figure{coarsening_2d.ps}{Growth of $L(t)$ during coarsening in two dimensions}
{18.0cm}{18.0cm} Figure 9

\Figure{coarsening_3d_loglog_new.ps}{Growth of $L(t)$
during coarsening in three dimensions: log--log plot}
{18.0cm}{18.0cm} Figure 10

\FigureComposite{comparison_of_configurations}
{Configurations during coarsening for $J_2 = 0$ and $J_2 \ne 0$}
{\topinsert}
{
\centerline{
{\psfig{figure=\figdir/slice_inf_36.ps,height=7.0cm,width=7.0cm}}
{\hskip .5cm}
{\psfig{figure=\figdir/slice_inf_120.ps,height=7.0cm,width=7.0cm}}
}
\centerline{\sectionfont (a) {\hskip 2.4cm} {$J_2 = 0$} {\hskip 2.4cm} {(b)}}
\bigskip\bigskip
\centerline{
{\psfig{figure=\figdir/slice_6_1000.ps,height=7.0cm,width=7.0cm}}
{\hskip .5cm}
{\psfig{figure=\figdir/slice_6_10000.ps,height=7.0cm,width=7.0cm}}
}
\centerline{\sectionfont (c) {\hskip 2.4cm} {$J_2 \ne 0$} {\hskip 2.4cm} {(d)}}
} Figure 11


\Figure{sample_tiling_new.ps}{A sample configuration of the tiling model}
{12.0cm}{12.0cm} Figure 12

\FigureIgnoreHeight{illegal_tiling.ps}{An illegal configuration in the tiling
model} {12.0cm}{12.0cm} Figure 13

\FigureComposite{tiling_dynamics}
{Elementary dynamical move in the tiling model}
{\topinsert}
{
\centerline{
{\psfig{figure=\figdir/tiling_before_flip.ps,width=7.5cm}}
{\hskip .5cm}
{\psfig{figure=\figdir/tiling_after_flip.ps,width=7.5cm}}
}
\centerline{\sectionfont (a) {\hskip 7.5cm} {(b)}}
} Figure 14

\FigureComposite{tilings_during_coarsening}
{Configurations during coarsening in the tiling model}
{\topinsert}
{
\centerline{
{\psfig{figure=\figdir/tiling_config_t10_4.ps,width=8.0cm}}
{\hskip .5cm}
{\psfig{figure=\figdir/tiling_config_t10_5.ps,width=8.0cm}}
}
\centerline{{\hskip 1.2cm} $t=10^4$,{\hskip 0.5em} $L=3.4$
	{\hskip 5.0cm} {$t=10^5$,{\hskip 0.5em} $L=4.6$}}
\bigskip
\centerline{
{\psfig{figure=\figdir/tiling_config_t10_6.ps,width=8.0cm}}
{\hskip .5cm}
{\psfig{figure=\figdir/tiling_config_t10_7.ps,width=8.0cm}}
}
\centerline{{\hskip 1.2cm} $t=10^6$,{\hskip 0.5em} $L=5.6$
	{\hskip 5.0cm} {$t=10^7$,{\hskip 0.5em} $L=7.0$}}
\bigskip
\centerline{
{\psfig{figure=\figdir/tiling_config_t10_8.ps,width=8.0cm}}
{\hskip .5cm}
{\psfig{figure=\figdir/tiling_config_t10_9.ps,width=8.0cm}}
}
\centerline{{\hskip 1.2cm} $t=10^8$,{\hskip 0.5em} $L=8.8$
	{\hskip 5.0cm} {$t=10^9$,{\hskip 0.5em} $L=9.9$}}
} Figure 15

\Figure{coarsening_tiling_loglog.ps}{Growth of $L(t)$
during coarsening in the tiling model: log--log plot}
{18.0cm}{18.0cm} Figure 16

\Figure{coarsening_tiling_longtime_new.ps}{Growth of $L(t)$
during coarsening in the tiling model: very late times}
{18.0cm}{18.0cm} Figure 17

\FigureComposite{tilings_thermal_fluctuations}
{Thermal fluctuations along domain boundaries}
{\topinsert}
{
\centerline{
{\psfig{figure=\figdir/tiling_config_t10_4.ps,width=14.5cm}}
}
\centerline{$T=3J_2$, {\hskip 0.7em} $L=3.4$}
\bigskip
\centerline{
{\psfig{figure=\figdir/tiling_config_T5_t10_4.ps,width=14.5cm}}
}
\centerline{$T=5J_2$, {\hskip 0.7em} $L=3.2$}
} Figure 18

\Figure{annealing_tiling.ps}{Growth of $L$
during cooling in the tiling model}
{18.0cm}{18.0cm} Figure 19

\Figure{final_annealing_tiling.ps}{Final value of L as
a function of the cooling rate $\Gamma$}
{18.0cm}{18.0cm} Figure 20


\FigureComposite{examples}
{Example of how to compute the energy of a configuration}
{\topinsert}
{
\centerline{
{\psfig{figure=\figdir/example_interface1.ps,width=6.5cm}}
{\hskip .5cm}
{\psfig{figure=\figdir/example_interface2.ps,width=6.5cm}}
}
\centerline{\sectionfont (a) {\hskip 6.0cm} (b)}
} Figure 21

\Figurepage{scaling_2d_new.ps}
{Scaling Plot for Correlation Function in 2D for $J_2 \ne 0$}{18.0cm}{18.0cm}
Figure 22

\Figurepage{scaling_3d_new.ps}
{Scaling Plot for Correlation Function in 3D for $J_2 \ne 0$}{18.0cm}{18.0cm}
Figure 23

\endpaper

\end